\pdfoutput=1
\documentclass[cernpreprint,texlive=2016,txfonts,UKenglish]{latex/atlasdoc}
\usepackage[subfigure]{latex/atlaspackage}
\usepackage{latex/atlasbiblatex}
\usepackage{latex/atlascontribute}
\usepackage{latex/atlasphysics}
\graphicspath{{logos/}{figures/}}
\AtlasTitle{\Large Search for top quark decays $t\rightarrow qH$, with $H\to\gamma\gamma$, in $\sqrt{s} =$ 13 \tev\ $pp$ collisions using the ATLAS detector}

\author{The ATLAS Collaboration}

\PreprintIdNumber{CERN-EP-2017-118}

\AtlasJournalRef{JHEP10(2017)129}
\AtlasDOI{10.1007/JHEP10(2017)129}

\AtlasJournal{JHEP}
\AtlasAbstract{%
This article presents a  search for flavour-changing neutral currents in the decay of a top quark into an
up-type ($q = c, u$) quark and a Higgs boson, where the Higgs boson decays into two photons. The proton--proton
collision data set analysed amounts to 36.1 fb$^{-1}$ at $\sqrt{s} = 13\ \tev$ collected by the ATLAS 
experiment at the LHC. Top quark pair events are searched for, where one top quark decays into $qH$ and the
other decays into $bW$. Both the hadronic and leptonic decay modes of the $W$ boson are used. No significant
excess is observed and an upper limit is set on the $t \rightarrow cH$ branching ratio of $2.2\times 10^{-3}$ at
the 95\% confidence level, while the expected limit in the absence of signal is $1.6 \times 10^{-3}$. The
corresponding limit on the $tcH$ coupling is 0.090 at the 95\% confidence level. The observed upper limit on
the $t \rightarrow uH$ branching ratio is $2.4 \times 10^{-3}$.
}

\hypersetup{pdftitle={ATLAS document},pdfauthor={The ATLAS Collaboration}}

\def\nttb{$t\bar{t}$}
\def\nbbb{$b\bar{b}$}
\def\nffb{$f\bar{f}$}
\def\mgg{\hbox{$m_{\gamma\gamma}$}}
\makeatletter
\DeclareOldFontCommand{\rm}{\normalfont\rmfamily}{\mathrm}
\DeclareOldFontCommand{\bf}{\normalfont\bfseries}{\mathbf}
\DeclareOldFontCommand{\it}{\normalfont\itshape}{\mathit}
\makeatother

\begin{document}

\maketitle

\tableofcontents

\newpage
\section{Introduction}
\label{sec:introduction} 
The discovery of the Higgs boson in 2012, by the ATLAS and CMS collaborations~\cite{ObservationATLAS,ObservationCMS},
has opened the possibility of searching, among the abundant \nttb\ final states  produced in proton--proton collisions at the CERN Large Hadron Collider 
(LHC),  for the decay  $t\rightarrow qH$ of a top quark (or antiquark) into  an on-shell Higgs boson, of mass 125.09~\gev ~\cite{Aad:2015zhl}, plus an up-type quark (or antiquark), $q=c$ or $u$. This decay, which proceeds via a flavour-changing neutral current (FCNC), is forbidden at tree level in the Standard Model (SM) and, with respect to the dominant top quark decay mode ($t\rightarrow bW$), is suppressed at higher orders due to the Glashow--Iliopoulos--Maiani mechanism~\cite{Glashow:1970gm}. The  $t\rightarrow qH$ branching ratio (${\cal B}$) in the SM is estimated to be around $3\times 10^{-15}$ (see Ref.~\cite{AguilarSaavedra:2004wm} and references therein), and an observation of this decay would provide a clear signal of physics beyond the SM (BSM).

\par In BSM models,
different mechanisms can yield effective couplings orders of magnitude larger than those of the SM.
Examples of such extensions are the quark-singlet 
model~\cite{AguilarSaavedra:2002ns, delAguila:1998tp, 
AguilarSaavedra:2002kr}, the two-Higgs-doublet model with or 
without flavour violation~\cite{cheng:1987rs, Grzadkowski:1990sm, 
Luke:1993cy, Atwood:1995ud, Atwood:1996vj, Bejar:2000ub,Baum:2008, Chen:2013, Atwood:2013}, 
the minimal supersymmetric standard model~\cite{Li:1993mg, deDivitiis:1997sh, Lopez:1997xv, 
Guasch:1999jp, Delepine:2004hr, Liu:2004qw, Cao:2007dk}, Supersymmetry with R-parity 
violation~\cite{Yang:1997dk}, the Topcolour-assisted Technicolour 
model~\cite{Lu:2003yr}, models with warped extra 
dimensions~\cite{Agashe:2004cp,Agashe:2006wa, Agashe:2009di} and the Littlest Higgs model with T-parity
conservation~\cite{2013:LHT}. In composite Higgs boson models, FCNC may appear even with a single Higgs doublet~\cite{Agashe:2009di}, inducing $t\rightarrow cH$ branching ratios up to $10^{-4}$.
For a recent review, see Ref.~\cite{Agashe:2013hma}. 

\par Among all the possibilities, the largest branching ratio ${\cal B} \sim 1.5\times 10^{-3}$ appears in the ansatz of Cheng and Sher~\cite{cheng:1987rs}
in which the tree-level coupling remains allowed in the absence of an additional symmetry, 
and scales with the top and up-type quark masses, $m_t$ and $m_q$, as 
$\lambda_{tqH}=\sqrt{2m_qm_t}/v$ (where $v = 246\ \gev$ is the Higgs field vacuum expectation value). This can be regarded as the ``maximal naturalness'' approach in the Higgs flavour structure. 
\par In this article, the $t\rightarrow qH$ decay is searched for using 36.1 \ifb\ of 13 \tev\ proton--proton collision data taken in 2015 and 2016 with the ATLAS detector at the LHC. The search uses  the $H\rightarrow\gamma\gamma$ decay mode, taking advantage of the clean diphoton signature, despite its small branching ratio (0.23\% in the SM~\cite{deFlorian:2016spz}).
\par A search for $t\rightarrow qH$ decays was previously performed at $\sqrt s$ = 7 and 8 \tev\ with Run-1 data, both in ATLAS and in CMS. As the $t\rightarrow cH$ and $t\rightarrow uH$ modes are hard to distinguish, the upper limits are given assuming that each final state is produced alone. In ATLAS, the $H\rightarrow\gamma\gamma$ channel~\cite{Aad:2014dya}  yields the observed and expected limits on ${\cal B}(t\rightarrow cH)$ of 0.79\% and 0.51\% respectively, at the 95\% confidence level (CL). Combining with the  multi-lepton\footnote{The multi-lepton final state includes events from the $H\to ZZ^{*}, WW^{*}$ and  $\tau^+\tau^-$ decays.} and $H \to$ \nbbb\ final states, which, at 8 \tev, are each comparable in sensitivity to the $H\rightarrow\gamma\gamma$ channel,
the overall observed (expected) limit at the 95\% CL is 0.46\% (0.25\%) for $t\rightarrow cH$ and 0.45\% (0.29\%) for $t\rightarrow uH$ ~\cite{Aad:2015pja}.
The 95\% CL observed (expected) limits from the CMS Collaboration on the  $t\rightarrow cH$ and $t\rightarrow uH$ branching ratios are 0.40\% (0.43\%) and 0.55\% (0.40\%) respectively~\cite{Khachatryan:2016atv}, using the same final states as ATLAS.

Compared to the previous ATLAS result in the $H\rightarrow \gamma\gamma$ mode, a significant
 improvement is attained due to the large increase (about four times) in the \nttb\ production cross section when increasing the centre-of-mass energy from 8 \tev\ to 13 \tev, and the larger integrated luminosity recorded during 2015 and 2016. Adverse effects are the increase in the production cross section of backgrounds, and the significantly larger ``pile-up'' noise due to the superimposition of additional interactions on the selected hard interaction.
 \par The analysis aims to select \nttb\ pairs with one top quark decaying into $bW$ (SM decay) and the other into $qH$. It is split into two final states, targeting the decay of the $W$ boson from the SM top quark decay either in the hadronic mode (hadronic selection) or in the leptonic mode (leptonic selection). The result is extracted from a fit to the diphoton invariant mass spectra of a resonant signal function centred around the Higgs boson mass and a background function, mainly constrained by bands on either side of the signal regions. While this data driven approach does not require a detailed understanding of the background, the dominant contributions are however presented together with the data, at different stages of the analysis. 
\par In this analysis, SM production of the Higgs boson, with its subsequent decay into a diphoton final state, is considered as a background, described by the same diphoton invariant mass distribution as the sought-for FCNC production. Its magnitude, determined by simulation, is accounted for in the signal extraction.

\section{Detector, data set and simulation samples}
\label{sec:Detector and Data}
\subsection{ATLAS detector}
\par The ATLAS detector~\cite{ATLAS_detector} consists of an inner detector for tracking
surrounded by a superconducting solenoid providing a 2\,T magnetic field,
electromagnetic and hadronic calorimeters, and a muon spectrometer. The inner detector provides tracking in the pseudorapidity
\footnote{ATLAS uses a right-handed coordinate system with its origin at the nominal interaction point
(IP) in the centre of the detector and the $z$-axis along the beam line. Observables labelled as transverse are projected onto the $x$--$y$ plane. The
$x$-axis points from the IP to the centre of the LHC ring, and the $y$-axis
points upwards. Cylindrical coordinates $(r, \phi)$ are used in the transverse
plane, $\phi$ being the azimuthal angle around the beam line. The pseudorapidity
is defined in terms of the polar angle $\theta$ as $\eta = -\ln\tan(\theta/2)$. The angular distance $\Delta R$ is defined as
$\Delta R \equiv \sqrt{(\Delta \eta)^2 + (\Delta \phi)^2}$. The transverse energy is $\et = E/\cosh(\eta)$.}
 region $|\eta| < 2.5$ and consists of silicon pixel and microstrip detectors
 inside a transition radiation tracker that covers $|\eta| < 2.0$.  A new innermost silicon pixel layer has been added to the inner detector after the Run-1 data taking~\cite{ibl}.
The electromagnetic calorimeter, a lead/liquid-argon sampling device with accordion geometry, is divided into
 one barrel ($|\eta|<1.475$) and two end-cap ($1.375<|\eta|<3.2$) sections.
 Longitudinally, it is divided into three layers. While most of the energy is deposited in the second layer, the first layer, referred to as
 the strip layer, has fine segmentation in the regions $|\eta|<1.4$ and $1.5<|\eta|<2.4$ to facilitate the separation
 of photons from neutral hadrons and to allow shower
 directions to be measured. In the range of $\lvert\eta\rvert < 1.8$, a
 presampler layer allows the energy to be corrected for losses
 upstream of the calorimeter. The barrel ($|\eta|<1.7$) hadronic calorimeter consists
 of steel and scintillator tiles, while the end-cap sections ($1.5<|\eta|<3.2$)
 are composed of copper and liquid argon. The forward calorimeter
 ($3.1<|\eta|<4.9$) uses copper and tungsten as absorber with liquid argon as active material.
The muon spectrometer consists of precision ($|\eta|<2.7$) and trigger ($|\eta|<2.4$) chambers equipping a toroidal magnet system which surrounds the hadronic calorimeter.
\subsection{Data set}
\label{subsec:dataset}
\par This analysis uses the full proton--proton data set recorded by ATLAS in 2015 and 2016 with the LHC operating at a centre-of-mass energy $\sqrt{s} = 13$ \tev\ and a bunch spacing of 25 ns. After application of data-quality requirements, the integrated luminosity amounts to 36.1 \ifb, with a relative uncertainty of 3.2\%.\footnote{The uncertainty is derived, following a methodology similar to that detailed in Ref.~\cite{Aaboud:2016hhf}, from a preliminary calibration of the luminosity scale using $x$--$y$ beam-separation scans performed in August 2015 and May 2016.}
The data were recorded with instantaneous luminosities up to $1.4\times 10^{34}\,\text{cm}^{-2}\text{s}^{-1}$.
The mean number of interactions per bunch crossing,
$\mu$, was on average 13 in 2015 
and 25 in 2016. 
The inelastic collisions that occur in addition to the hard interaction produce mainly low transverse momentum particles that form the pile-up background.

\par The data considered here were selected using a diphoton trigger which requires two clusters formed from energy deposits in the electromagnetic calorimeter. The transverse energy threshold was 35 \gev\ (25 \gev) for the leading (sub-leading) cluster (sorted in $E_\mathrm{T}$). Loose criteria were applied to ensure that the shape of selected clusters matched that expected for electromagnetic showers initiated by photons. 
The efficiency of the trigger for events containing two photons passing the  offline selection requirements of this analysis was measured to be
larger than 99\%. 

\subsection{Signal and background simulation}
\label{subsec:simulation}
The FCNC signal was simulated using \textsc{MG5\_aMC@NLO 2.4.3}~\cite{amcatnlo} interfaced to  \textsc{Pythia 8.212}~\cite{Sjostrand:2007gs}
with the \texttt{A14}~\cite{a14} set of tuned parameters for the modelling of parton showers, hadronisation and multiple interactions.
The \nttb\ pairs were generated at next-to-leading order (NLO) in quantum chromodynamics (QCD) with the TopFCNC UFO model~\cite{Degrande:2014tta}, using the 5-flavour scheme
and the NNPDF3.0~\cite{nnpdf30} parton distribution functions (PDFs). The factorisation and renormalisation scales were set equal to
$\sqrt{m_t^2+(p_{\mathrm{T},t}^2+p_{\mathrm{T},\bar{t}}^2)/2}$, where $p_{\mathrm{T},t}\ (p_{\mathrm{T},\bar{t}})$ is the transverse momentum of the top quark (antiquark). The top quark decay was performed by \textsc{MadSpin}, and the Higgs boson decay by \textsc{Pythia 8}. Two samples corresponding to \nttb\ production
with one top quark decaying into a charm quark and a Higgs boson (which itself decays into two photons 
with 100\% branching ratio) were produced. The two samples, added together, correspond to the $W$ bosons from $t\to bW$ decaying leptonically or hadronically. The
leptonic decays of the $W$ include all three lepton flavours. The top quark mass taken in the simulation
is 172.5~\gev, and the Higgs boson mass is 125~\gev. Two equivalent samples with $t\rightarrow uH$ were also produced.
The \nttb\ cross section used is $832^{+40}_{-46}$ pb. It has been calculated at next-to-next-to leading order in QCD including resummation of next-to-next-to-leading logarithmic soft gluon terms with Top++2.0~\cite{Czakon:2013goa,Czakon:2011xx}.

\par The contribution from known production sources of Higgs bosons, followed by a decay into two photons, is considered here as a resonant background. 
The following production modes, ordered by their cross sections, were considered and simulated~\cite{Hgaga13}: gluon--gluon fusion ($ggH$), vector-boson 
fusion (VBF) and associated production ($WH$, $ZH$, \nttb$H$, $b\bar{b}H$, $tHjb$, and $tWH$). 
The Higgs boson cross sections and branching ratios compiled by the LHC Higgs Cross-Section Working Group~\cite{deFlorian:2016spz} are used for normalisation. For
$ggH$, the QCD next-to-next-to-next-to-leading-order cross section is used. For $tWH$, the NLO normalisation given in Ref.~\cite{Demartin:2016axk} is used.
	
The $ggH$ and VBF events were generated at NLO using \textsc{Powheg}~\cite{Alioli:2008tz,Nason:2009ai} interfaced  to \textsc{Pythia 8} with the \texttt{AZNLO}~\cite{aznlo} set of tuned parameters for parton showering, hadronisation and multiple interactions. 
The events from $WH$ and $ZH$ associated production were generated at leading order with \textsc{Pythia 8}.
Associated \nttb$H$ production was generated at NLO with \textsc{MG5\_aMC@NLO} interfaced with \textsc{Pythia 8}. 
	The  parton distribution functions CT10~\cite{PDF:CT10}, NNPDF3.0 and CTEQ6L1~\cite{PDF:CTEQ6L1} were used for the \textsc{Powheg}, 
 \textsc{MG5\_aMC@NLO} and \textsc{Pythia 8} samples, respectively.
\par An event sample corresponding to non-resonant diphoton production, labelled S$_{\gamma\gamma j}$ in the following,  was  generated with the \textsc{Sherpa 2.1.1} event generator~\cite{Gleisberg:2008ta}, with up to three additional partons in the matrix element. This sample is used
as a benchmark sample for non-resonant background production. 
\par Samples of events from \nttb, $W$ and $Z$ production with one or two photons generated in the matrix element were also simulated. Due to initial-state radiation (ISR), final-state radiation (FSR) and jets faking photons, the one- and two-photon samples partly overlap,
the first one including an approximation of the second one. Given the cross sections of the simulated processes, and the statistical size of the available samples, the \nttb$\gamma$, $W \gamma\gamma$ and $Z \gamma\gamma$ samples were selected to represent each of the backgrounds, albeit with a large uncertainty in the corresponding cross sections. 
\par 
	The stable particles, defined as particles with a lifetime longer
	than 10~ps, were passed through a full detector
	simulation~\cite{atlasSim} based on \textsc{Geant4}~\cite{geant4}.
	The resulting ``particle hits'' in the active detector material were
	subsequently transformed into detector signals during digitisation.
\par Pile-up was modelled using simulated minimum-bias events generated using \textsc{Pythia 8}.
	The number of  events overlaid onto the hard-scattering events during the digitisation was randomly chosen so as to reproduce the distribution
of $\mu$ observed in data. 
	The effects from pile-up events occurring in nearby bunch crossings (out-of-time pile-up) were also modelled.

\section{Event reconstruction and candidate selection}
\subsection{Event reconstruction}
\label{subsec:offline}
\par While the requirement of two tightly identified and precisely measured photons is the key part of the selection, the analysis also requires jets and $b$-tagging for the hadronic selection as well as identified electrons or muons and missing transverse momentum (with magnitude \met) for the leptonic selection. A  minimum angular distance ($\Delta R$) is required between these objects. If some objects overlap after reconstruction, a removal is performed, keeping, in order of priority, photons, then leptons, and finally jets. 
\par The photon reconstruction~\cite{Aaboud:2016yuq} is seeded by clusters of energy deposits in the
electromagnetic calorimeter with transverse energy greater than 2.5~\gev\ in a region of $0.075\times0.125$ in $\eta\times\phi$. The accepted $|\eta|$ region extends from 0 up to 2.37 excluding the region [1.37, 1.52], which is less instrumented and where dead material affects both the identification and the energy measurement. 
The reconstruction is designed to separate electrons from unconverted and converted photons. 
Clusters without any matching track or conversion vertex are classified as
unconverted photon candidates while clusters with a matching conversion vertex are classified as converted photon candidates. In simulation, when requiring 
the generated photon \et to be above 20~\gev, the reconstruction efficiency is 98\% on average for converted and unconverted photons.
\par The photon energy is measured from clusters of size $0.075\times 0.175$ in $\eta\times\phi$ in the barrel region of the calorimeter
and $0.125\times 0.125$ in the calorimeter end-caps, using a combination of simulation-based 
and data-driven calibration factors~\cite{Aad:2014nim} determined from $Z\rightarrow e^{+} e^{-}$ events collected in 2015 and 2016. The photon energy resolution in simulation is corrected to match the resolution in data~\cite{calib2015}. 
\par The identification of photons~\cite{Aaboud:2016yuq} is based on lateral and longitudinal shower shapes measured in the
electromagnetic calorimeter. Two working points, ``loose'' and ``tight'', are defined. Photon candidates are 
required to deposit only a small fraction of their energy in the hadronic calorimeter and to have a lateral
shower shape consistent with that expected from a single electromagnetic shower. 
The information about the shape of the shower from the highly segmented strip layer of the calorimeter is used to
discriminate single photons from hadronic jets in which a neutral meson carries most of the jet energy. The ``tight'' identification efficiency, averaged over $\eta$, for unconverted (converted) photons ranges from 85\% to 95\% (90\% to 98\%) for \et\ between 25 \gev\ and 200 \gev~\cite{photon2015}.
\par To suppress the hadronic background, the photon candidates are required to be isolated from any
other significant activity in the calorimeter and the tracking detectors. The calorimeter isolation is
computed as the sum of the  transverse energy of positive-energy topological clusters\footnote{Topological clusters are three-dimensional
clusters of variable size, built by clustering calorimeter cells on the basis
of the signal-to-noise ratio~\cite{TopoClusters}.} reconstructed in a cone of  $\Delta R = 0.2$ around 
the photon candidate. The transverse energy of the photon candidate is subtracted from the sum. The contributions of 
underlying event and pile-up are subtracted based on the method suggested in Ref.~\cite{jetarea}.  
The track isolation is computed as the scalar sum of the transverse momentum \pt\ of all tracks with $\pt > 1\ \gev$ within  a cone of $\Delta R = 0.2$  around  the photon candidate. The tracks must satisfy some loose track quality criteria and originate from
the interaction vertex (see below). For converted photon candidates, the tracks associated with the
conversion are removed. For the ``loose'' isolation working point chosen in this analysis, the calorimeter (track) isolation is required 
to be less than 6.5\% (5\%) of the photon transverse energy. The efficiency of the calorimeter isolation requirement ranges from 83\% to 97\% 
for photons of \et\ between 25 \gev\ and 200 \gev\ which pass the identification requirements.
The efficiency of the track isolation requirement ranges from 96\% to 100\% (93\% to 97\%)
for unconverted (converted) photons of \et\ between 25 \gev\ and 200 \gev\ which pass the identification and calorimeter
isolation requirements. 
A dedicated study using electrons and positrons from $Z$ boson decays as a proxy 
for photons showed that the ratio of  isolation efficiencies in data and simulation  remains constant at $1.01 \pm 0.01$ per event, for events with up to four reconstructed jets.  
 
\par The identification of the interaction vertex uses a neural network algorithm based on tracks and primary vertex information, as well as on the two photon directions measured in the calorimeter
and inner detector (in the case of a conversion)~\cite{Aad:2014eha}. Given the presence of jets in the selected sample, the vertex identification efficiency is high, with 97\% of selected vertices falling within $\pm$ 1 mm of the true position, for events with four or more reconstructed jets. 

\par Jets are reconstructed from topological clusters using the anti-$k_t$ algorithm~\cite{JetAlgo} with a radius parameter $R=0.4$ and are required
to have an $|\eta_{\rm det}|<4.4$, where $\eta_{\rm det}$ is the jet pseudorapidity assuming the interaction point is at the nominal collision 
position.
The dependence of the jet response on the number of primary vertices and the average number of interactions is mitigated 
by applying a data-driven event-by-event subtraction procedure based on the jet area~\cite{jetarea}.
In order to suppress jets produced in additional pile-up interactions, each jet with $|\eta|<2.4$ and  
$\pt <60\ \gev$ is  required to have a ``jet vertex tagger (JVT) parameter'' value larger than 0.59~\cite{Aad:2015ina}.
The corresponding efficiency for jets from the hard interaction, in a sample of simulated $ggH$ events, with $H \to \gamma\gamma$,  
is estimated to be 86\%, with a purity (fraction of jets originating from the hard interaction)
of 90\%. These fractions are significantly higher (94\% and 99\% respectively), for the sample of simulated \nttb $H$ events, with $H \to \gamma\gamma$.
\par Jets containing $b$-hadrons are identified ($b$-tagged) using the MV2c10 tagger~\cite{mv2,Aad:2015ydr}, with a selection
threshold corresponding to a 77\% efficiency 
for jets containing a $b$-hadron in \nttb\ events. The corresponding rejection factor of jets originating from light quarks or gluons is $\sim 130$
and it is $\sim 6$ for charm quark jets. 

\par Electron candidates consist of clusters of energy deposited in the electromagnetic 
calorimeter that are associated with tracks in the inner detector~\cite{ElectronPerformance}, and are consistent with those expected for 
electromagnetic showers. The electron 
transverse momentum is computed from the cluster energy and the track direction at the 
interaction point. Furthermore, all tracks associated with 
electromagnetic clusters are refitted using a Gaussian-sum Filter~\cite{GSF}, which 
accounts for bremsstrahlung energy losses. Electron candidates are required to satisfy $\et > 15\ \gev$ and $|\eta_{\rm cl}|< 2.47$, excluding  $1.37<|\eta_{\rm cl}|< 1.52$ ($\eta_{\rm cl}$ is the pseudorapidity of the electromagnetic cluster). In addition, the transverse impact parameter
significance of the associated track must be below $5 \sigma$ and the longitudinal impact parameter must fulfil
$|z_{0}\sin\theta|<0.5$ mm. Electron candidates are required to be of ``medium'' quality~\cite{isoEl}, and to satisfy loose calorimeter
and track isolation criteria 
with thresholds tuned to have an average efficiency of $\sim 98\%$. 

\par Muons are required to meet the conditions $|\eta|< 2.7$ and $\pt > 10$ \gev, and to be of ``medium'' quality~\cite{isoMu}.
The transverse impact parameter significance of the muon tracks must be below $3 \sigma$ and the longitudinal impact parameter 
must fulfil $|z_{0}\sin\theta|<0.5$ mm. 
Muon candidates are required to satisfy track isolation~\cite{isoMu}  criteria with a \pt independent efficiency of 99\%.

\par Missing transverse momentum is calculated as the negative vector sum of the transverse momentum of all objects in an event~\cite{Aad:2016nrq}.  
Jets not satisfying the JVT criterion or with $\pt < 20\ \gev$ are removed, but their associated tracks originating from the interaction vertex are included in a dedicated {\it soft term}.

 The selection of  candidate events starts by 
applying the ``tight diphoton selection'': at least two photons satisfying the ``tight'' identification criteria, 
with loose calorimeter and track isolation, $\pt > 40\ \gev$ ($\pt > 30\ \gev$) for the leading (sub-leading) photon candidate, and diphoton invariant mass between 100 \gev\ and 160 \gev. The number of events passing this requirement in data is about four hundred 
thousand. Events without 
identified leptons (electron or muon) enter the hadronic selection; those with exactly one lepton enter the leptonic selection. Events with two or more identified leptons are rejected.
\subsection{Hadronic selection}
\label{subsec:FullHadronic}
\par Jets passing the reconstruction and selection described in Section~\ref{subsec:offline} are used together with 
the two photons to select the final sample. The significance of the FCNC signal was optimised with $S/\sqrt{B}$ as figure of merit,
where $S$ and $B$ are the numbers of events from the $t\rightarrow cH$ simulated signal sample and the S$_{\gamma\gamma j}$ sample,
respectively, both evaluated in the signal region (see below).
\par To reduce the contribution of pile-up jets, only jets with \pt $>30$ \gev\ are considered.
While in the background sample the distribution of the number of jets is a rapidly falling distribution, its counterpart
 for signal shows a broad peak around three to four. Events with at least four jets are kept for the hadronic
 analysis (about 9000 events remain at this stage), and only
 the four leading jets are used further in the analysis. In addition, it is required that at least one of the four jets is $b$-tagged.

\par The next step is designed to select events for which the six-body final state (two photons and four jets) is compatible with a \nttb\  final state. It starts by forming three-body objects: the two photons plus one jet on one side (Top1), and the three other jets on the other side (Top2). By grouping each of the four jets with the two photons, four (Top1,Top2) pairs are constructed, with corresponding invariant masses ($M_{1}$,$M_{2}$). For an event to be selected, there must be at least one combination (Top1,Top2) with masses ($M_{1}$,$M_{2}$) compatible with the top quark mass, as decribed below.\footnote{An additional condition that the $b$-tagged jet belongs to Top2 was evaluated but not retained, as this slightly worsens the sensitivity for the $t\rightarrow c H$ mode, where the charm quark contributes to the overall $b$-tagging efficiency of an event.}
\par The distributions of $M_{1}$ and $M_{2}$ are shown in Figure~\ref{hadronic:fig:Mt12BT}; the  $M_{1}$ distribution has four entries per event, while for  $M_{2}$ only those combinations passing the  $M_{1}$ selection, as described below, enter. In the signal distributions, the peaks from correctly reconstructed top quark decays are clearly visible, and particularly narrow in the Top1 distribution, while the combinatorial background has a shape similar to the distribution obtained with the S$_{\gamma\gamma j}$ sample. The S$_{\gamma\gamma j}$ sample is normalised to data, after subtraction of the \nttb$\gamma$ contribution. The $W\gamma\gamma$ and $Z\gamma\gamma$ contributions are negligible.
 \par Based on the position and width of the two signal peaks,
the window chosen for the $M_{1}$ selection ranges from 152 \gev\ to 190 \gev, while for $M_{2}$ the broader range from 120 \gev\ to 220  \gev\ is chosen. 
\begin{figure}[h!]
\begin{center}
\subfigure[]{\includegraphics[width=.45\textwidth]{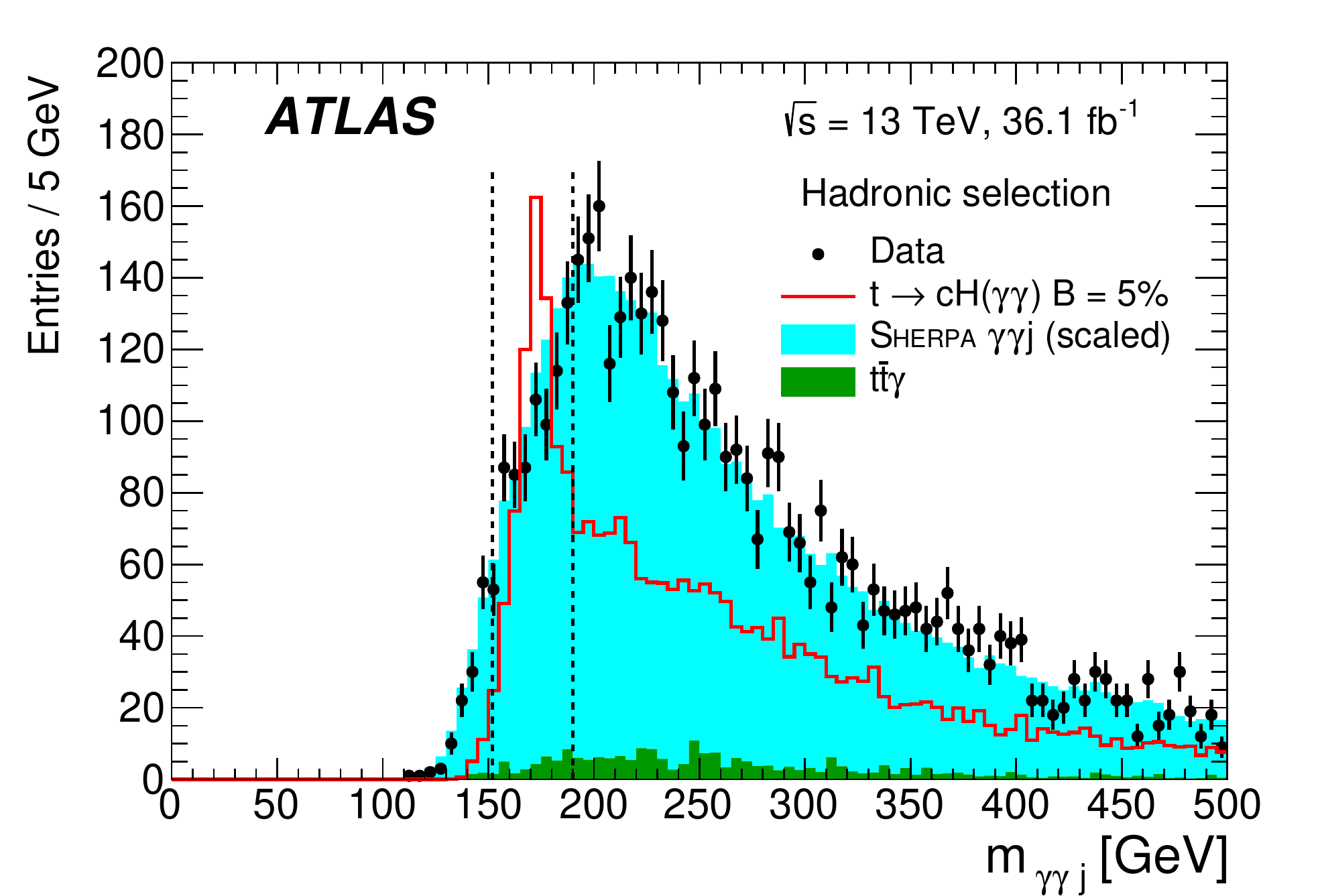}}
\subfigure[]{\includegraphics[width=.45\textwidth]{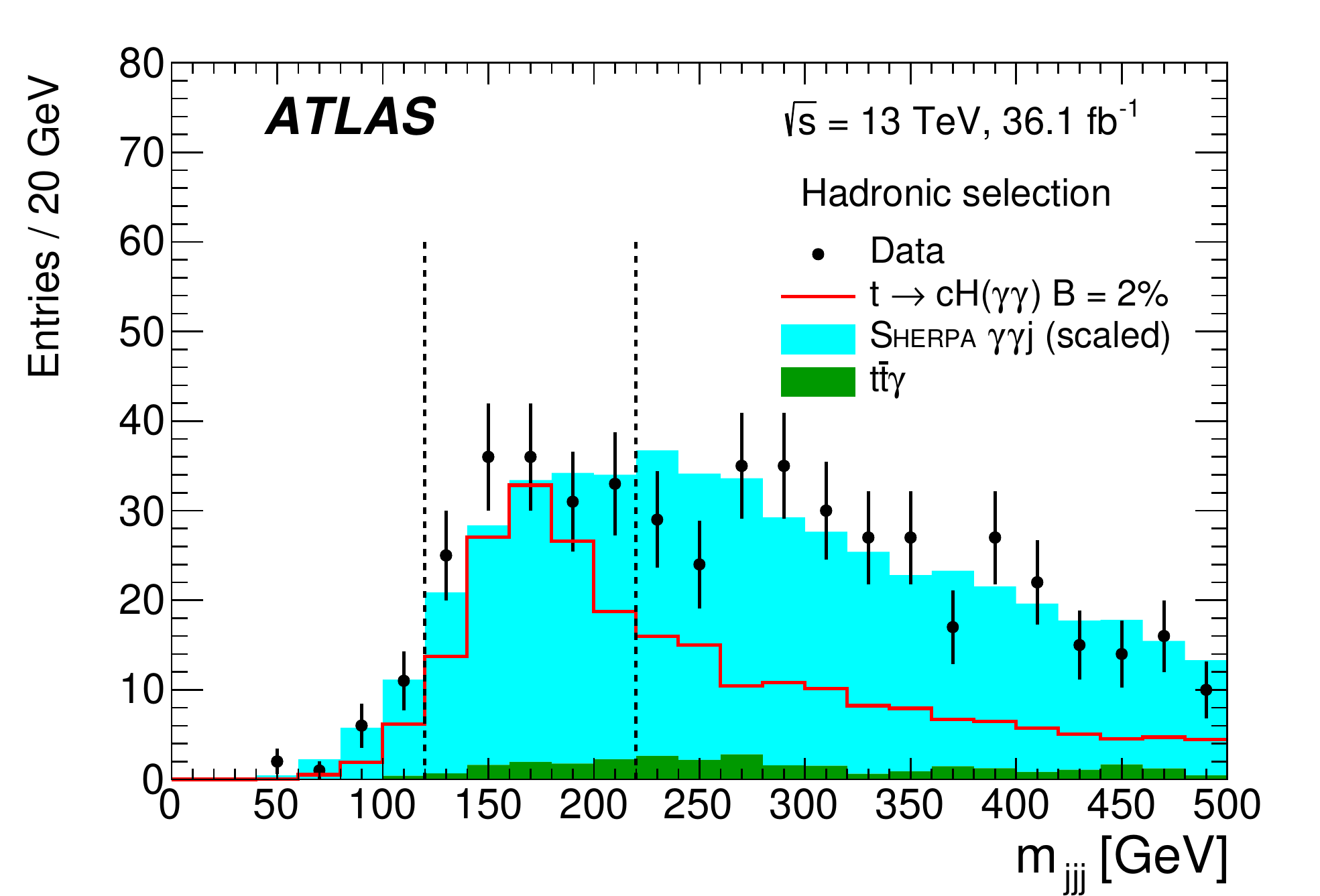}}
\end{center}
\vspace{-0.5cm}
  \caption{Distribution of the invariant mass of (a) the two selected photons, together with one of the four selected 
jets (Top1, four entries per event), and (b) the three remaining jets (Top2, one entry per combination fulfilling the $M_{1}$ selection). At least one $b$-tagged jet per event is required. The \nttb$\gamma$ distributions, normalised to the data's integrated luminosity using the theoretical cross section, are superimposed, together with the \textsc{Sherpa} S$_{\gamma\gamma j}$ contribution normalised to the difference between data and \nttb$\gamma$. The signal distributions are normalised assuming a branching ratio of (a) 5\% and (b) 2\%. The vertical dotted lines indicate the (a) $M_{1}$ and (b) $M_{2}$ selection windows (see text).}
  \label{hadronic:fig:Mt12BT}
\end{figure}

\par In order to increase the acceptance, albeit with a reduced signal-to-noise ratio, events failing the $M_{2}$ selection step are also retained for the final analysis by exploiting two (orthogonal) categories:
\begin{itemize}
\item category 1: events that pass the full selection;
\item category 2: events that fail the $M_{2}$ requirement but satisfy all other selection criteria.
\end{itemize}
The category 1 (2) data sample has 115 (437) events.  
The corresponding acceptances for the simulated \nttb\ events with $t\rightarrow cH$ and $H\rightarrow\gamma\gamma$ are $(2.89\pm0.10)\%$ and $(4.15\pm0.12)\%$ for category 1 and 2 respectively, where the uncertainties are statistical only. During the optimisation phase of the analysis,  the region between 122 \gev\ and 129 \gev\ where the sought-for FCNC signal is expected to appear, and called the signal region (SR) in the following, was masked (with a margin of 2 \gev\ on the low side and 1 \gev\ in the high side). The category 1 (2) data sample  has 14 (69) events with the diphoton mass falling in the signal region.

 \par The $\gamma\gamma$ invariant mass spectra for data and for the S$_{\gamma\gamma j}$ and \nttb$\gamma$ samples are compared in 
Figures~\ref{hadronic:fig:Mgg_cat12}(a) and \ref{hadronic:fig:Mgg_cat12}(b) for categories 1 and 2, respectively. The normalisation of simulation to data excludes the mass range from 122 to 129 \gev\ in both cases. The \nttb$\gamma$ contribution is normalised to the integrated luminosity of the data using its theoretical cross section, while S$_{\gamma\gamma j}$  covers the difference between data and \nttb$\gamma$.
The expected SM Higgs boson production and an additional FCNC signal  normalised assuming ${\cal B}=0.2\%$ are shown stacked over the sum of the backgrounds.

\par Good agreement is observed between the shapes of the data and the $\gamma\gamma$+jets background represented by the S$_{\gamma\gamma j}$ sample, for the $M_{1}$  and $M_{2}$ spectra shown in Figure~\ref{hadronic:fig:Mt12BT}.  Since this background and the data agree rather well in the sidebands of the $\gamma\gamma$ mass spectra (Figure~\ref{hadronic:fig:Mgg_cat12}), the S$_{\gamma\gamma j}$ sample is used (see Section~\ref{sec:StatandRes}) to determine the background fitting function needed for the final signal extraction.
\begin{figure}[h!]
\begin{center}
\subfigure[]{\includegraphics[width=.45\textwidth]{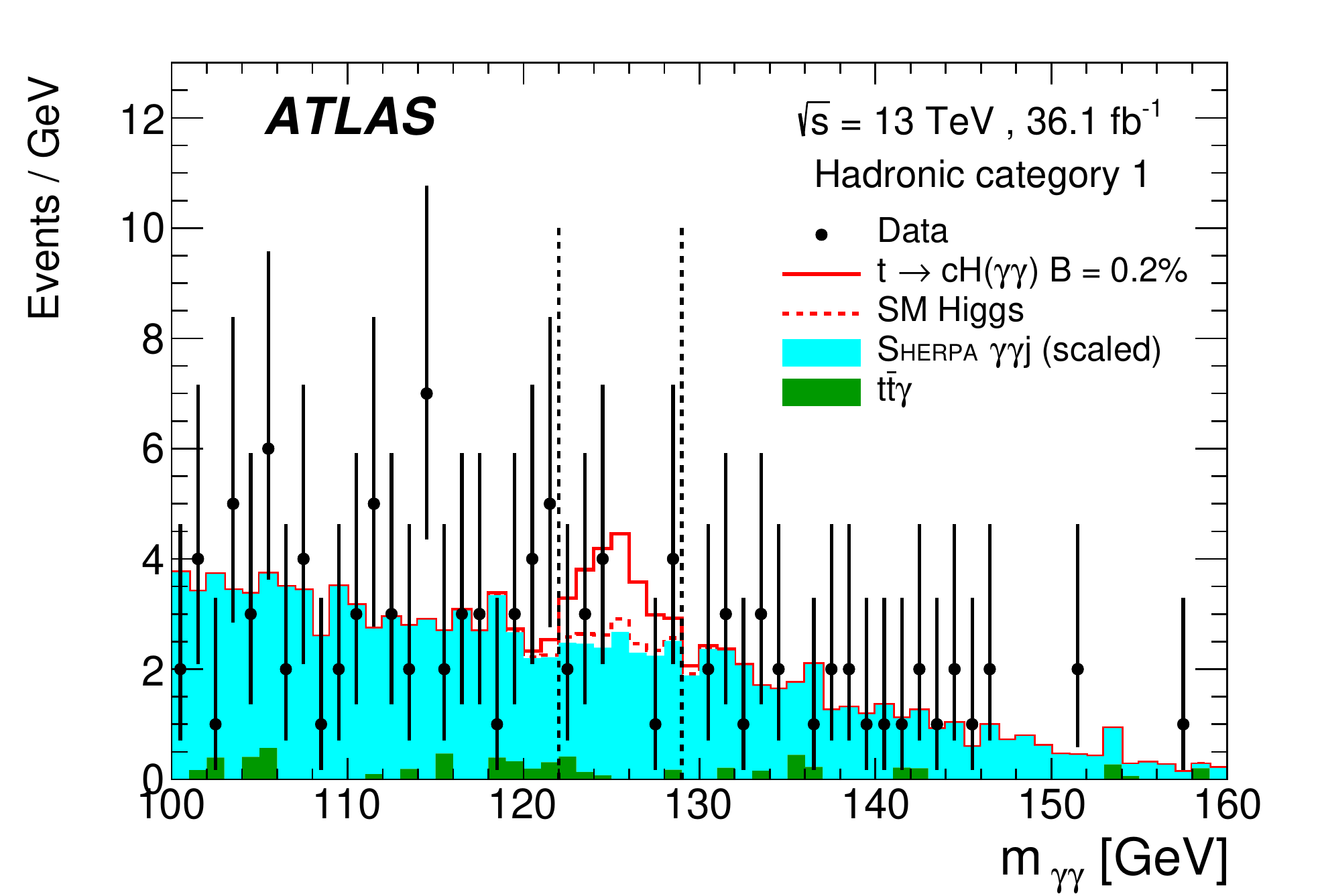}}
\subfigure[]{\includegraphics[width=.45\textwidth]{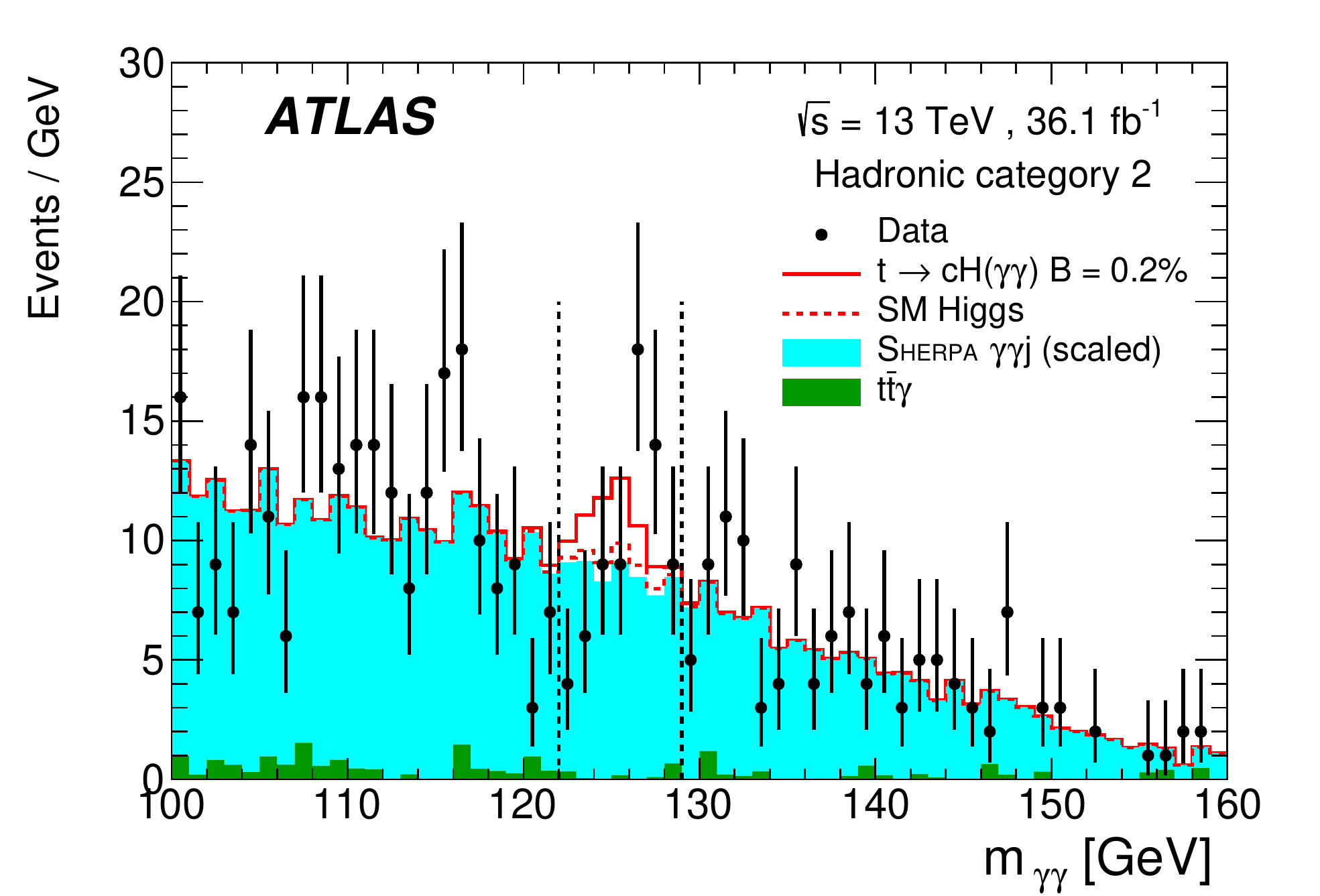}}
\end{center}
\vspace{-0.5cm}
  \caption{Distributions of the diphoton invariant mass, for data events, the \nttb$\gamma$ sample, and the S$_{\gamma\gamma j}$ sample for (a) category 1 and (b) category 2 of the hadronic selection. The $t\bar{t}\gamma$ sample is normalised using the data's integrated luminosity and the theoretical cross section. The \textsc{Sherpa} S$_{\gamma\gamma j}$ sample is normalised to the difference between data and $t\bar{t}\gamma$. The SM Higgs boson production and an additional FCNC signal normalised assuming ${\cal B} = 0.2\%$ are shown stacked over the sum of the backgrounds. The vertical dotted lines indicate the signal region (see text).}
  \label{hadronic:fig:Mgg_cat12}
\end{figure}

\subsection{Leptonic selection}
\label{subsec:Leptonic}
The aim of the leptonic analysis is to identify \nttb\ events where one top quark decays into $qH$, and the $W$ boson originating from the other top quark decays leptonically.
Only electrons and muons are considered as candidates for lepton identification.

\par The selection of leptonic candidate events starts by applying the ``tight diphoton selection''. Requiring in addition one identified
 lepton  reduces the sample to 833 events. The \pt threshold for leptons is set to 10 \gev\ for muons, and 15 \gev\ for electrons.
 The higher threshold in the electron case is motivated by  a comparatively larger fraction of jets faking electrons. 
Requiring two or more jets with \pt $>30$ \gev\ further reduces the sample to 223  events. Only the two leading jets are used further in the analysis.
\par The dominant background sources are \nttb$\gamma$, $W\gamma\gamma$ and $Z\gamma\gamma$ and the diphoton + 
jets background, in which a small fraction of jets are wrongly identified as leptons.

\par In order to further select leptonic signal events, the missing transverse
momentum is used to determine, with the lepton \pt, the transverse mass $m_\mathrm{T}$ whose distribution is shown in
Figure~\ref{leptonic:fig:mtrans}(a) for data and the relevant simulated samples. In the region between 56 \gev\ and 88 
\gev\ the 
background prediction, dominated by \nttb$\gamma$, $W\gamma\gamma$ and  $Z\gamma\gamma$, significantly underestimates the data: while $25\pm3\%$ of data events fall in this region, the corresponding fraction for the simulation is $16\pm1\%$, where the uncertainties
are statistical only. The discrepancy is attributed to the large uncertainty affecting the theoretical cross sections of these
 processes and to the lack of $W\gamma$ and $Z\gamma$ events with an extra photon from ISR, FSR or jets faking photons, for which the
 size of the simulated samples was not large enough to have a meaningful estimate after the full selection. Applying a scaling factor of 3
 to the $W\gamma\gamma$ and $Z\gamma\gamma$ contributions (as derived from the comparison of the \nttb$\gamma$ and \nttb$\gamma\gamma$ samples), the fraction of simulated events with $m_\mathrm{T}$ between 56 \gev\ and 88 \gev\ becomes $27\pm1\%$, thus compatible with the data. As explained in Section~\ref{sec:introduction}, the disagreement as observed in Figure~\ref{leptonic:fig:mtrans}(a) does not affect the result because in the final fit
 (see Section~\ref{sec:StatandRes}) the background is determined from data only. 
A selection requiring $m_\mathrm{T}$ larger than 30 \gev, which preserves a large fraction of the signal and rejects much of the background (especially from the  S$_{\gamma\gamma j}$  sample), is applied. The number of data events remaining at this stage of the analysis is 124.
\begin{figure}[h!]
\begin{center}
\subfigure[]{ \includegraphics[width=.45\textwidth]{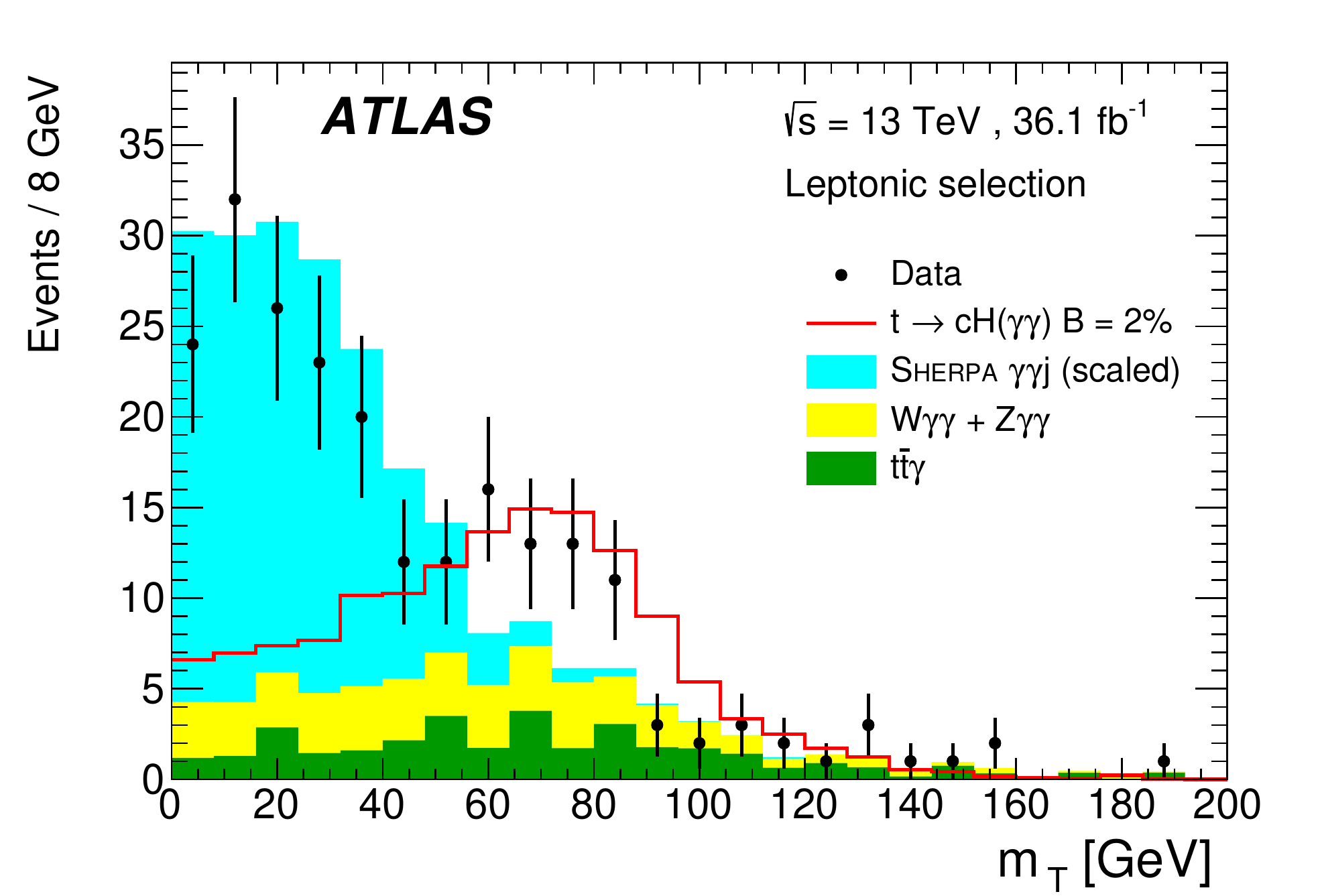}}
\subfigure[]{ \includegraphics[width=.45\textwidth]{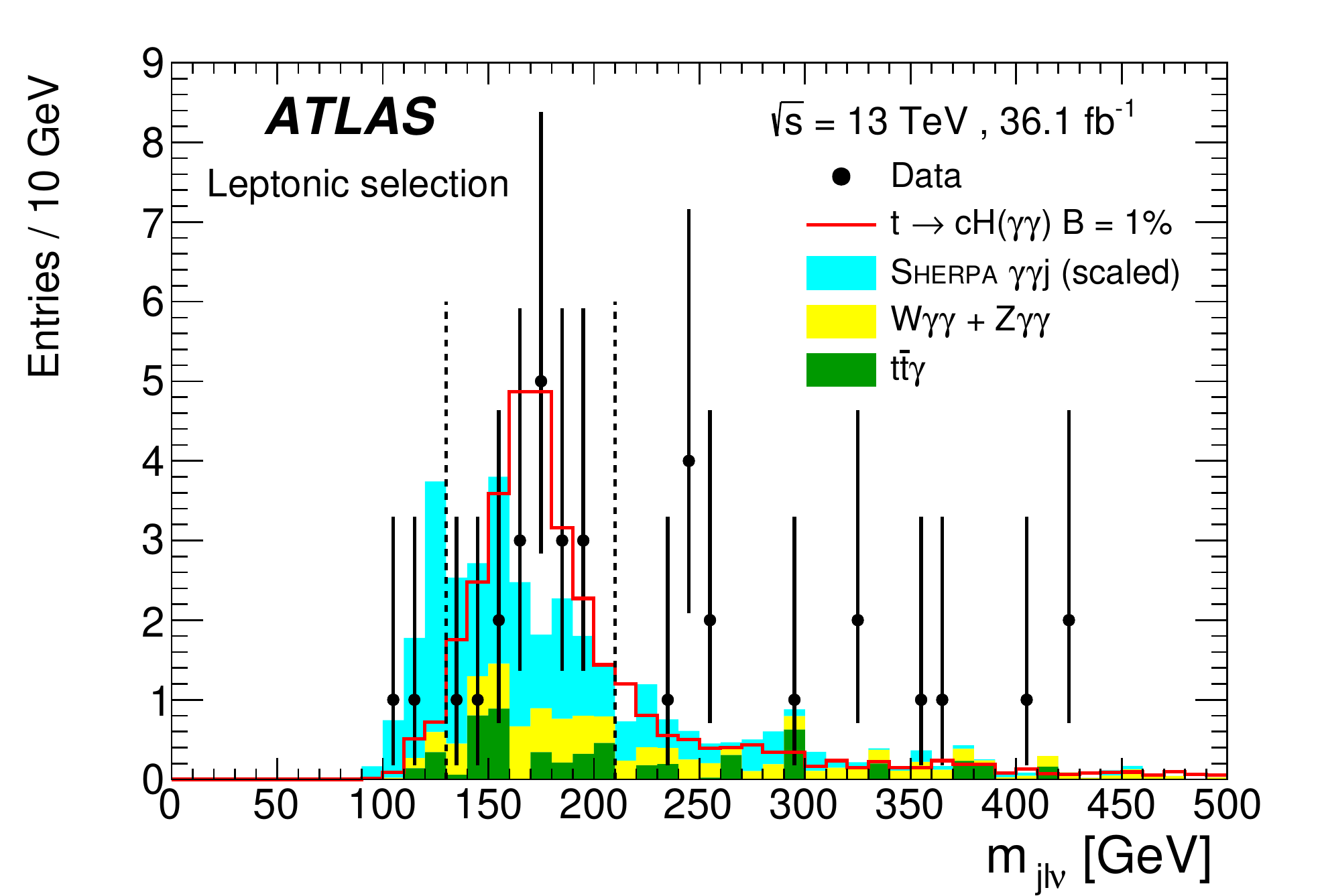}}
  \hspace{-0.5cm}
\end{center}
\vspace{-0.5cm}
  \caption{Distribution (a) of the transverse mass calculated from the lepton kinematics and the missing transverse momentum and (b) of the invariant mass of the lepton, the neutrino, and one jet for each $\gamma\gamma j$ combination where the $m_{\gamma\gamma j}$ mass falls in the $M_1$ acceptance window. No $b$-tagging is required. The \nttb$\gamma$, $W\gamma\gamma$ and $Z\gamma\gamma$ distributions are superimposed, normalised to the data's integrated luminosity using theoretical cross sections. The \textsc{Sherpa} S$_{\gamma\gamma j}$ sample is normalised to the difference between data and the sum of \nttb$\gamma$, $W\gamma\gamma$ and $Z\gamma\gamma$. The distribution of the FCNC signal is normalised assuming a branching ratio of (a) 2\% and (b) 1\%. The vertical dotted lines in (b) indicate the $M_{2}$  selection window (see text).}
  \label{leptonic:fig:mtrans}
\end{figure}

\par The next step is to verify, as was done for the hadronic selection, that the final-state particles are  kinematically compatible with the decay of two top quarks. The invariant mass $M_{1}$ of the two photons and one of the two jets (Top1) is calculated, as well as the mass $M_{2}$ of the remaining jet, the lepton, and the neutrino (Top2). For the latter, the neutrino longitudinal momentum is estimated by using a $W$ boson mass constraint, as was done in Ref.~\cite{Aad:2014dya}.
The same calculation is repeated, exchanging the role of the two jets. If the invariant masses ($M_{1}$, $M_{2}$) of one of the two (Top1, Top2) combinations 
fall in predefined windows around the top quark mass, the event is selected, provided one of the two jets is $b$-tagged. This defines category-1 events. 
Events fulfilling all requirements, except the one on $M_{2}$ are kept as category-2 events. As was done for the hadronic mode, the acceptance windows were optimised, resulting in the same interval for  $M_{1}$  (152 \gev\ to 190 \gev), and in a slightly narrower interval for  $M_{2}$, from 130 \gev\ to 210 \gev.
  
\par The event sample remaining at the end of the selection  is extremely small. Three events are selected in each of category 1 and category 2. The corresponding acceptances for the simulated \nttb\ events with $t\rightarrow cH$ and $H\rightarrow\gamma\gamma$ are $(0.96\pm0.03)\%$ for category 1 and $(0.27\pm0.02)\%$ for category 2, where the uncertainties are statistical only.  The corresponding diphoton invariant mass distributions are shown in Figure~\ref{leptonic:fig:Mgg_cat12}. In this figure, the \nttb$\gamma$, $W\gamma\gamma$ and $ Z\gamma\gamma$ backgrounds, affected by large statistical fluctuations, are shown using large bins, 10 \gev\ wide. Only \nttb$\gamma$ remains significant at this stage. Its contribution is normalised to the data's integrated luminosity using the theoretical  cross section. Like for the hadronic analysis, the  contributions from S$_{\gamma\gamma j}$  are normalised to match the number of events observed in the sidebands, once the \nttb$\gamma$ background is taken into account. No contribution of this type is actually necessary for category 1.
The expected SM Higgs boson contribution and an additional FCNC signal normalised assuming ${\cal{B}}=0.2\%$ are shown stacked over the backgrounds.

 \begin{figure}[h!]
\begin{center}
\subfigure[]{\includegraphics[width=.45\textwidth]{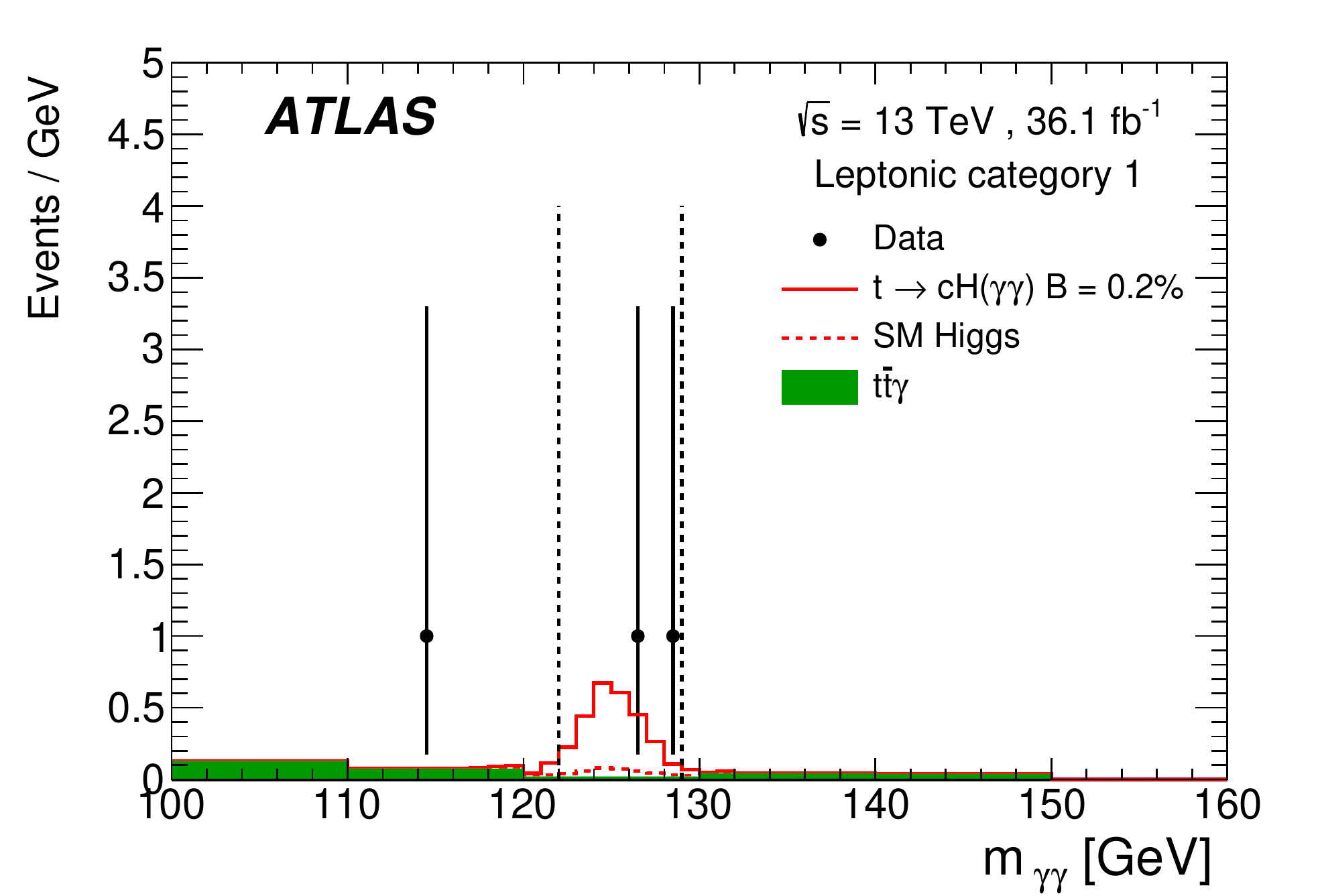}}
\subfigure[]{\includegraphics[width=.45\textwidth]{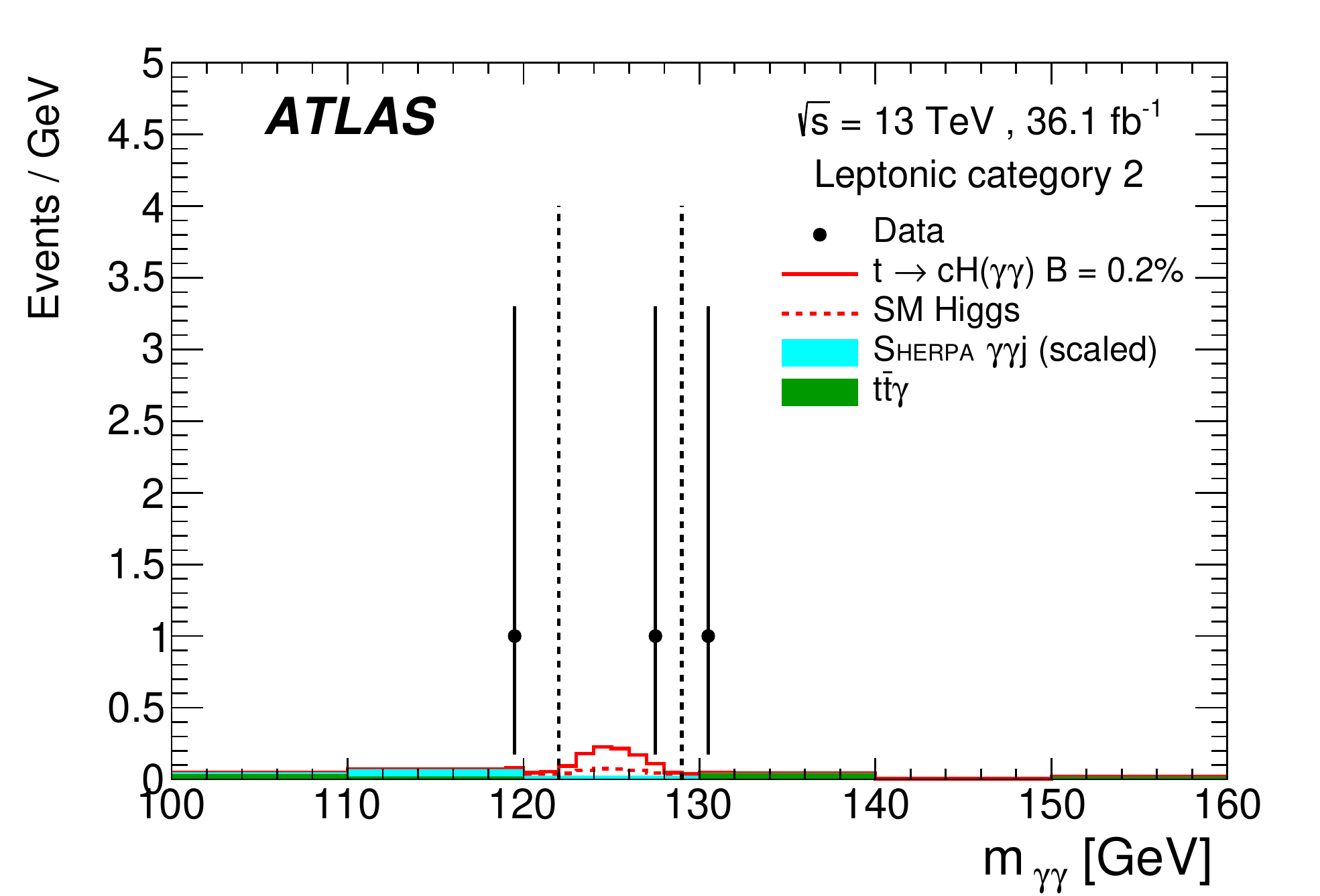}}
\end{center}
\vspace{-.5cm}
 \caption{Distributions of the diphoton invariant mass for (a) category 1 and (b) category 2 of the leptonic analysis.  The \nttb$\gamma$ distribution is normalised to the data's integrated luminosity using the theoretical cross section. The \textsc{Sherpa} S$_{\gamma\gamma j}$ contribution is shown for category 2, where its contribution is significant (see text). The SM Higgs boson production and an additional FCNC signal  normalised assuming ${\cal{B}}$ = 0.2\% are shown stacked over the backgrounds. The vertical dotted lines indicate the signal region (see text).} 
  \label{leptonic:fig:Mgg_cat12}
\end{figure}

\section{Statistical analysis and systematic uncertainties }
\label{sec:StatandRes}
The branching ratio ${\cal B}$ of the decay $t \rightarrow c(u)H$ is determined in a fit to data by using
a likelihood function, $L$, which is the product of the likelihoods for the four
channels (hadronic and  leptonic selections, each with two categories). 
Hypothesised values of ${\cal B}$ are probed with a test statistic based on the profile likelihood ratio~\cite{asym}. 
For the hadronic selection a fit to the $m_{\gamma\gamma}$ distributions in category 1 and category 2 is performed. The analysis in the leptonic selection is  based on event counting, for each category, in two $m_{\gamma\gamma}$ regions: the control region (CR), from 100 to 122 \gev\ and from 129 to 160 \gev, and the signal region.
\par The theoretical uncertainties are mainly related to the \nttb\ production cross section, the Higgs boson branching ratio to $\gamma\gamma$, the
 resonant background from SM processes and the signal generator uncertainties. The experimental and the generator systematic uncertainties are detailed in Section~\ref{subsec:Expsys}. All of them are introduced as nuisance parameters in the likelihood.
\subsection{Expected event yields}
\label{sec:normsignalandSM}
The relevant acceptances and expected resulting numbers of events for the $t \to cH$ signal simulation are
reported in Table~\ref{tab:Sighadlep}  for the hadronic and leptonic analyses, taking 0.2\% as branching ratio for $t \rightarrow cH$.
\par A fraction of the SM Higgs boson events produced in association with jets (and one lepton) is accepted by the hadronic
(leptonic) analysis if the Higgs boson decays into two photons. These events  appear as an additional resonant contribution
to the sought-for FCNC diphoton signal.
All known SM processes, as listed in Section~\ref{subsec:simulation}, were simulated to obtain the acceptances for the hadronic and leptonic selections. Summing all channels gives the expected numbers of events listed in Table~\ref{tab:Sighadlep}. The \nttb$H$ production mode gives the dominant contribution for both selections and both categories. The uncertainties include statistical uncertainties and uncertainties in the cross sections, as given in Ref.~\cite{deFlorian:2016spz}.
\par Dalitz decays (of the type $H \rightarrow \gamma$\nffb, where $f$ is a light fermion) are present in both the signal and SM Higgs boson simulations while the 0.23\% branching ratio taken for the $H\rightarrow\gamma\gamma$ decay does not include these processes. As their acceptance is essentially zero, a consistent normalisation is restored by removing their contribution to the normalisation factor of each simulated sample. No associated uncertainty is assigned. 

\begin{table}[h!]
\begin{center}
\caption{Expected acceptances (including efficiencies) for simulated \nttb\ events, with $t \rightarrow cH$ and $H\rightarrow\gamma\gamma$. The 
corresponding expected numbers of events are also shown, together with those for SM Higgs boson production, followed by $H\rightarrow\gamma\gamma$. The fraction of events from the \nttb $H$ channel, dominant in the SM contribution, is also given. The uncertainties in the acceptances are only statistical. The uncertainties in the expected numbers of events include statistical uncertainties and theoretical uncertainties in the cross sections.} 
\label{tab:Sighadlep}
\footnotesize
\vspace*{1em}
\begin{tabular}{l|cc|cc}
  \hline \hline
Selection                       &  \multicolumn{2}{c|}{Hadronic} & \multicolumn{2}{c}{Leptonic} \\
Category                        & 1    & 2   & 1    & 2 \\ \hline 
                                & \multicolumn{4}{c}{Signal $t \rightarrow cH$} \\[0.8ex] 
Acceptance with stat. unc. [\%] & $2.89\pm0.10$     & \,\, $4.15\pm0.12$     & $0.96\pm0.03$      & $0.27\pm0.02$ \\[0.8ex]
Expected events for ${\cal B} = 0.2\%$   & $7.85^{+\mbox{\footnotesize 0.64}}_{-\mbox{\footnotesize 0.67}}$  & 
$11.30^{+\mbox{\footnotesize 0.91}}_{-\mbox{\footnotesize 0.96}}$  & $2.60^{+\mbox{\footnotesize 0.21}}_{-\mbox{\footnotesize 0.23}}$  & $0.71^{+\mbox{\footnotesize 0.07}}_{-\mbox{\footnotesize 0.07}}$ \\[0.8ex]
\hline 
                                & \multicolumn{4}{c}{SM Higgs boson resonant background} \\[0.8ex]
Expected events                          & $1.17^{+\mbox{\footnotesize 0.09}}_{-\mbox{\footnotesize 0.11}}$  & $3.27^{+\mbox{\footnotesize 0.25}}_{-\mbox{\footnotesize 0.27}}$   & $0.26^{+\mbox{\footnotesize 0.02}}_{-\mbox{\footnotesize 0.03}}$  & $0.23^{+\mbox{\footnotesize 0.02}}_{-\mbox{\footnotesize 0.02}}$   \\[0.8ex]
\nttb$H$ fraction               & 90\%               & 68\%                & 92\%               & 77\% \\ \hline \hline
\end{tabular}
\end{center}
\normalsize
\end{table}  
\subsection{Likelihood for the hadronic selection}
\label{sec:lkl}
The (unbinned) likelihood for each hadronic category includes a Poisson term, the product over all events of the expected $m_{\gamma\gamma}$ distribution function (background + signal parameterised as described below), and a term constraining the external parameters (nuisance parameters) to their expected values, within uncertainties, by the product of the corresponding Gaussian factors.
\par The signal distribution, assumed identical for SM Higgs boson production and the $t\rightarrow cH, H\rightarrow\gamma\gamma$ signal is described by a double-sided Crystal Ball function (a Gaussian function with power-law tails on both sides) whose parameters are obtained from a fit to the mass spectrum of the simulated FCNC signal. The Crystal Ball function's mean value is shifted by 90 MeV to account for the difference between the measured Higgs boson mass of 125.09 \gev, and the value of 125 \gev\ used in the signal simulation. The same signal parameterisation is used for the leptonic analysis (see below).

The parameterisations of the background are obtained from the $m_{\gamma\gamma}$ shapes of the S$_{\gamma\gamma j}$ sample.  In order to verify that the parameterisations
 do not induce a spurious signal when fitting the data, the two distributions are smoothed using non-parametric probability density
functions~\cite{Cranmer:2000du}, and used to generate background-only pseudo-experiments with on average 115 and 445
events for categories 1 and 2, respectively. The contribution from \nttb$\gamma$, which is small and has large
statistical fluctuations, is neglected. The corresponding $m_{\gamma\gamma}$ spectra are fitted with
different shape parameterisations. For background-only fits a bias is defined as the
difference between the true number of events and the fitted number of events in the SR. 
For signal+background fits, the bias is defined as the fitted number of signal events. The criterion used to select a background parameterisation
is that these biases have to be smaller than 10\% of the number of signal events at the expected limit or 20\% of the
expected uncertainty in this number, mainly resulting from the statistical uncertainty in the number of events in the SR. 
The background-only and signal+background fits give consistent results. Third- and fourth-order
polynomial parameterisations were found to satisfy this criterion and the one with fewer free parameters, the third-order polynomial,
was chosen. The associated biases are 0.5 and 1.7 events in category 1 and 2, respectively, and are used as a
systematic uncertainty in the final fit.

\subsection {Likelihood for the leptonic selection}
Given the very low observed number of events in each leptonic category, the likelihood for the leptonic channels is simply taken as a Poisson term, for two intervals of the diphoton invariant mass distribution, the SR and the CR. Instead of using a full probability density function, the distribution of the background is controlled by a free parameter $\alpha$, defined as the ratio of the numbers of background events expected in the SR and CR. The signal, $\sim 90\%$ of which is confined to the SR, is controlled by its magnitude, the other free parameter.
 The $\alpha$ parameter is one of the nuisance parameters, and its uncertainty is estimated by considering its variations as a function of the assumed background shapes. For a uniform \mgg\ distribution $\alpha = 0.13$ is expected. Using the smooth probability density function
from the hadronic analysis yields $\alpha \sim 0.17$.
The S$_{\gamma\gamma j}$ background remaining after the leptonic selection (see Section~\ref{subsec:Leptonic}), without any $b$-tagging requirement and summing categories 1 and 2 to enlarge the available 
sample,  gives $\alpha\sim 0.11$. The nominal value chosen in the fit is $\alpha = 0.14$, with a 30\% uncertainty.

\subsection{Systematic uncertainties} 
\label{subsec:Expsys}
A summary of the systematic uncertainties affecting the signal yields is given in Table~\ref{tab:Sumuncer}. The impact of these uncertainties (where relevant) on the parameters of the double-sided Crystal Ball functions is also taken into account. Since many of the uncertainties have a small impact and the
analysis is statistically limited, uncertainties affecting jet energy scale, $b$-tagging, \met, photon identification and lepton-related uncertainties
have been grouped so as to have only one  equivalent nuisance parameter for each of these quantities.

\begin{table}[h!]
\begin{center}
  \caption{Summary of theoretical, experimental and generator (see text) relative uncertainties in the signal yields, for the hadronic and leptonic selections (in percent, per event). The top three rows affect only the normalisation. The uncertainty related to the photon and lepton isolation selections is included in the identification uncertainty.
  }
\label{tab:Sumuncer}
\footnotesize
\vspace*{1em}
\begin{tabular}{l|cc|cc}
  \hline \hline
  & \multicolumn{4}{c}{Systematic uncertainty [\%]} \\
  \hline
Selection                &  \multicolumn{2}{c|}{Hadronic} & \multicolumn{2}{c}{Leptonic} \\ 
Category                 & 1    & 2   & 1    & 2 \\ \hline
\nttb\  production cross section & \multicolumn{4}{c}{ $^{+\mbox{\footnotesize 4.8}}_{-\mbox{\footnotesize 5.5}}$} \\[0.8ex]
Branching ratio of the $H\rightarrow\gamma\gamma$ decay & \multicolumn{4}{c}{\ \ 5.0} \\
Luminosity  & \multicolumn{4}{c}{\ \ 3.2} \\\hline
Photon energy scale      &\ \  0.3  & \ \ 0.5 &\ \  1.9  &\ \  0.6 \\
Photon energy resolution &\ \  0.3&\ \  0.2 &\ \  0.3  & \ \ 0.6 \\
Photon identification    &\ \  1.8  & \ \ 1.7 &\ \  1.7  &\ \  1.7     \\
Jet energy scale         & $^{+\mbox{\footnotesize 7.9}}_{-\mbox{\footnotesize 6.5}}$ & $^{+\mbox{\footnotesize 6.2}}_{-\mbox{\footnotesize 4.8}}$ & $^{+\mbox{\footnotesize 0.8}}_{-\mbox{\footnotesize 2.1}}$  & $^{+\mbox{\footnotesize 4.3}}_{-\mbox{\footnotesize 5.2}}$   \\[0.8ex]
Jet energy resolution    & $-2.1$ &\ \  3.0 & $-2.2$ &\ \  3.5               \\
Jet vertex tagging  & \ \ 0.7  & \ \ 0.4 & \ \ 0.3  & \ \ 0.2               \\
$b$-tagging              & \ \ 3.3  & \ \ 3.6 & \ \ 3.2  &\ \  6.5    \\
Lepton identification and scale    &  --  &  -- &\ \  0.6  &\ \  1.2         \\
\met\ scale              &  --  &  -- & $^{+\mbox{\footnotesize 1.1}}_{-\mbox{\footnotesize 0.6}}$ & $^{+\mbox{\footnotesize 1.8}}_{-\mbox{\footnotesize 3.9}}$ \\[0.8ex]
Pile-up reweighting      &\ \  1.6 & \ \ 1.2 &\ \  1.5  &\ \  1.2 \\ \hline
QCD scales               &\ \  1.7  & \ \ 1.6 &\ \  1.6  &\ \  3.2 \\
PDFs                     & \ \ 0.7  &\ \  0.5 &\ \  0.7  &\ \  0.5 \\ 
Top quark mass           &\ \  1.0  &\ \  1.0 & \ \ 1.0  &\ \  3.0 \\
Generator                & \ \ 5.7& $-5.8$ & 11.7 & \ \ 6.3\\
ISR/FSR                  & $^{+\mbox{\footnotesize 0.4}}_{-\mbox{\footnotesize 8.2}}$ & $^{-\mbox{\footnotesize 8.0}}_{+\mbox{\footnotesize 3.8}}$ & $^{-\mbox{\footnotesize 4.4}}_{+\mbox{\footnotesize 0.7}}$  & $^{-\mbox{\footnotesize 1.1}}_{+\mbox{\footnotesize 4.3}}$   \\[0.8ex]   
Hadronisation/underlying event       & $-5.8$  & \ \ 9.6& $-4.9$  & 23.2  \\ \hline \hline 
\end{tabular}
\end{center}
\end{table}
\begin{itemize}
\item The uncertainties in the \nttb\ production cross section, the $H \to \gamma\gamma$ branching ratio
  and the integrated luminosity, affect only the normalisation of the signal yield.
\item The systematic uncertainties in the energy scale ($\sim 0.5\%$) and resolution ($\sim 12\%$) of photons have
  a very small impact on the signal acceptance, as shown in Table~\ref{tab:Sumuncer}. They affect more significantly
  the parameters of the double-sided Crystal Ball functions used to fit the signal line shape. 
\item The uncertainties related to the photon trigger and identification amount to about 2\%, dominated by
the identification efficiency uncertainty.
\item The systematic uncertainty associated with the jet energy scale~\cite{Aaboud:2017jcu} is determined by changing each of the parameters to which it is sensitive by 
one standard deviation in each direction and one at a time, and taking the quadratic sum of all up (down) variations. The upper (lower) rows in the table correspond respectively to the up (down) variations.
 The same methodology is used for the jet energy resolution. The sign is negative if the yield decreases when the resolution improves. 
\item The differences in $b$-tagging efficiency between data and simulation are included in the event weights of the simulated 
samples. Replacing the corresponding scale factors by the values obtained when adding (subtracting) their uncertainty induces variations of the expected signal yield of the order of 3\%. The uncertainty related to the JVT selection is estimated in the same way.
\item The uncertainty associated with the lepton energy scale, identification and reconstruction efficiency, averaged for electrons and muons, is 
about 1\%. 
\item The uncertainty of about 2\% associated with \met\ was obtained with the same methodology as that used for the jet energy scale, 
  applied to the {\it soft term} as introduced in Section~\ref{subsec:offline}.
\item  The pile-up reweighting uncertainty accounts for the variations allowed when reweighting the distribution of the mean number of interactions per bunch crossing, $\mu$, from simulation to data. 
\end{itemize}
The uncertainties in the signal event generation are evaluated as follows:
\begin{itemize}
\item  The impact of the factorisation and renormalisation scales (QCD scales) on the signal acceptance is obtained by varying them by a factor of 0.5 or 2 with respect to the nominal values.
\item  The systematic uncertainty associated with the PDF choice is obtained by using the root mean square of the signal acceptance when considering the 100 Monte Carlo replicas available in the NNPDF3.0 set.
\item The systematic uncertainty related to the top quark mass uncertainty is obtained from the variation of acceptance observed when reweighting the simulated signal events by the ratio of a Breit--Wigner function at a mass $M$ (between 171 \gev\ and 174 \gev) to a Breit--Wigner
 function at a mass of 172.5 \gev\ (the mass used at event generation). 
The uncertainty in the Higgs boson mass, now known at $\pm 0.2\%$~\cite{Aad:2015zhl}, has a negligible impact on the result.
\item The uncertainty associated with the hard-process generation is obtained by comparing, at particle level~\footnote{ In a particle level simulation, the same reconstruction algorithms as for the full simulation are used (see Sections~\ref{subsec:simulation} and~\ref{subsec:offline}), but with final state particles as input instead of tracks and clusters.}, the acceptances obtained with
 \textsc{MG5\_aMC@NLO}  and \textsc{Powheg} for the \nttb\ generation (both interfaced to \textsc{Pythia 8}). The sign of the uncertainty is positive when \textsc{Powheg} gives a larger yield.
\item The uncertainty labelled ``ISR/FSR'' corresponds to the variation of the signal acceptance 
observed at particle level when the parameters governing QCD initial- and final-state radiation in \textsc{Pythia 8} are varied within the allowed 
range~\cite{a14}. The upper (lower) row in the table corresponds to the up (down) variation of both ISR and FSR.  
\item The systematic uncertainty associated with the hadronisation and underlying-event modelling is estimated  by comparing the acceptances of events
generated with \textsc{MG5\_aMC@NLO} interfaced to either \textsc{Pythia 8} or \textsc{Herwig~7}~\cite{Bellm:2015jjp}, at particle level. The sign of the uncertainty is positive when the yield from \textsc{Herwig~7} is larger.
\end{itemize}  
Finally the uncertainty in the normalisation of the expected yield for the SM Higgs boson production, as a resonant background, is obtained 
by combining the uncertainties in the individual cross-section predictions (from the renormalisation and factorisation scales, the PDFs and 
the strong coupling constant, as they are used for Table~\ref{tab:Sighadlep}), increased to
15\% for \nttb$H$, 100\% for $ggH$, VBF and $WH$, and 50\% for the other channels, to account for the fact that the simulations used
are not very accurate in the phase space probed by this analysis.

\section{Results}
\par Figure~\ref{fig:2dnll} shows the evolution of $q_{\cal B} = -2\left(\ln L({\cal B})-\ln L({\hat{\cal B}})\right)$ as a function of ${\cal B}$ for each category individually, and for the combined likelihood. The fitted branching ratio is ${\hat{\cal B}} = 6.9^{+6.8}_{-5.2}(\mathrm{stat.})^{+3.1}_{-1.5}(\mathrm{syst.})~\times~10^{-4}$. A summary of the
fitted numbers of signal and background events, together with the numbers of observed events in the SR, is given in Table~\ref{tab:summary}.

\begin{figure}[h!]
\begin{center}
    \includegraphics[width=.5\textwidth]{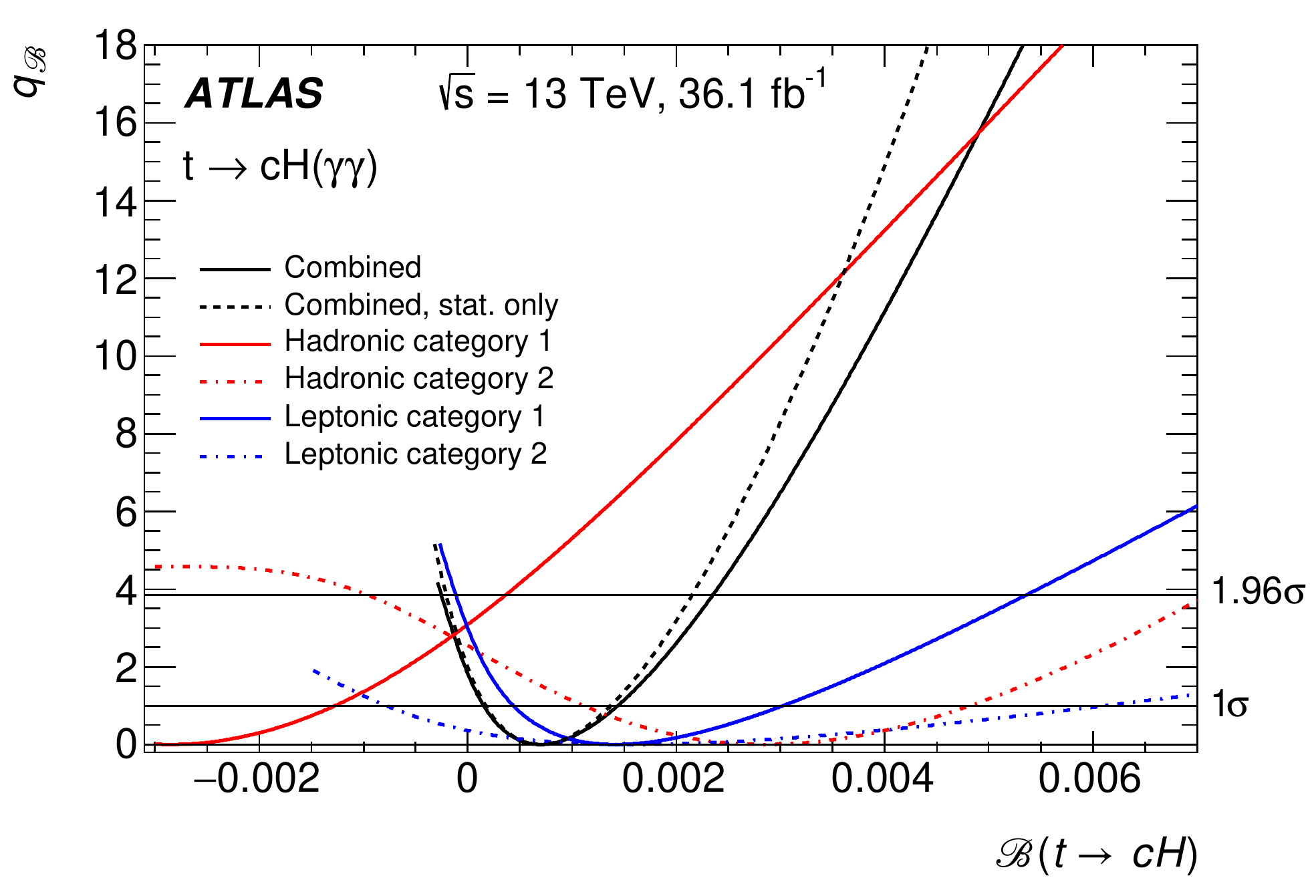}
\end{center}
\vspace{-0.5cm}
\caption{Observed evolution of $q_{\cal B}$ (see text) as a function of the $t\to cH$ branching ratio ${\cal B}$. The dotted curve corresponds to the case where all
  constrained nuisance parameters are fixed to their maximum-likelihood estimators at ${\hat{\cal B}}$. The likelihood functions are
  only defined for a positive number of expected events, hence the leptonic categories and combined curves do not cover the full scanned
  range. The intersection of the combined $q_{\cal B}$ curve with the line at $1.96 \sigma$ gives the 95\% CL upper limit in the asymptotic approximation.}
\label{fig:2dnll}
\end{figure}
\begin{table}[h!]
\begin{center}
\caption{Numbers of FCNC signal events in the SR ($m_{\gamma\gamma} \in [122,129]\ \gev$) for the fitted $t \to cH$ branching ratio ${\cal B} = 6.9\times10^{-4}$. The numbers of events for the SM Higgs boson production
  and the fitted non-resonant background are also shown, together with the number of observed events in data, in the four categories.}
\vspace{0.3cm}
\begin{tabular}{l|cc|cc}
\hline \hline
Selection     &  \multicolumn{2}{c|}{Hadronic} & \multicolumn{2}{c}{Leptonic} \\ 
Category                           & 1     & 2     & 1    & 2 \\ \hline 
Signal $t \rightarrow cH$          &\ \    2.4  &\   3.7  & 0.82  & 0.23   \\\hline 
SM Higgs boson resonant background &\ \   1.1  &\   3.1  & 0.24  & 0.22  \\ 
Other background                   & 16  & 63  & 0.14  & 0.29  \\
Total background                   & 17  & 66 & 0.38 & 0.51  \\ \hline 
Data                               & 14    & 69    & 2    & 1 \\ \hline \hline 
\end{tabular}
\label{tab:summary}
\vspace*{1em}
\end{center}
\normalsize
\end{table}  
\begin{figure}[h!]
\begin{center}
  \subfigure[]{\includegraphics[width=.45\textwidth]{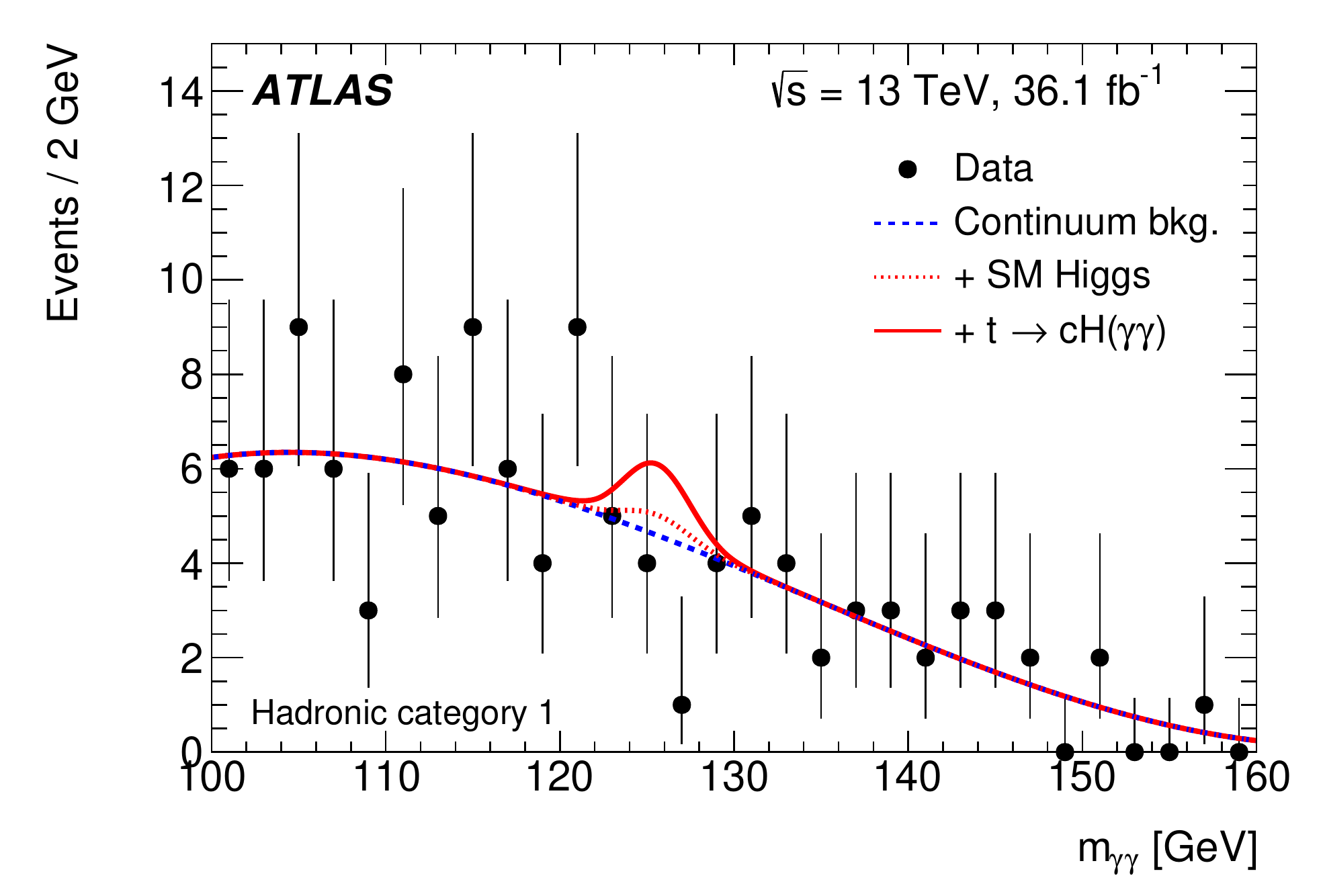}}
  \subfigure[]{\includegraphics[width=.45\textwidth]{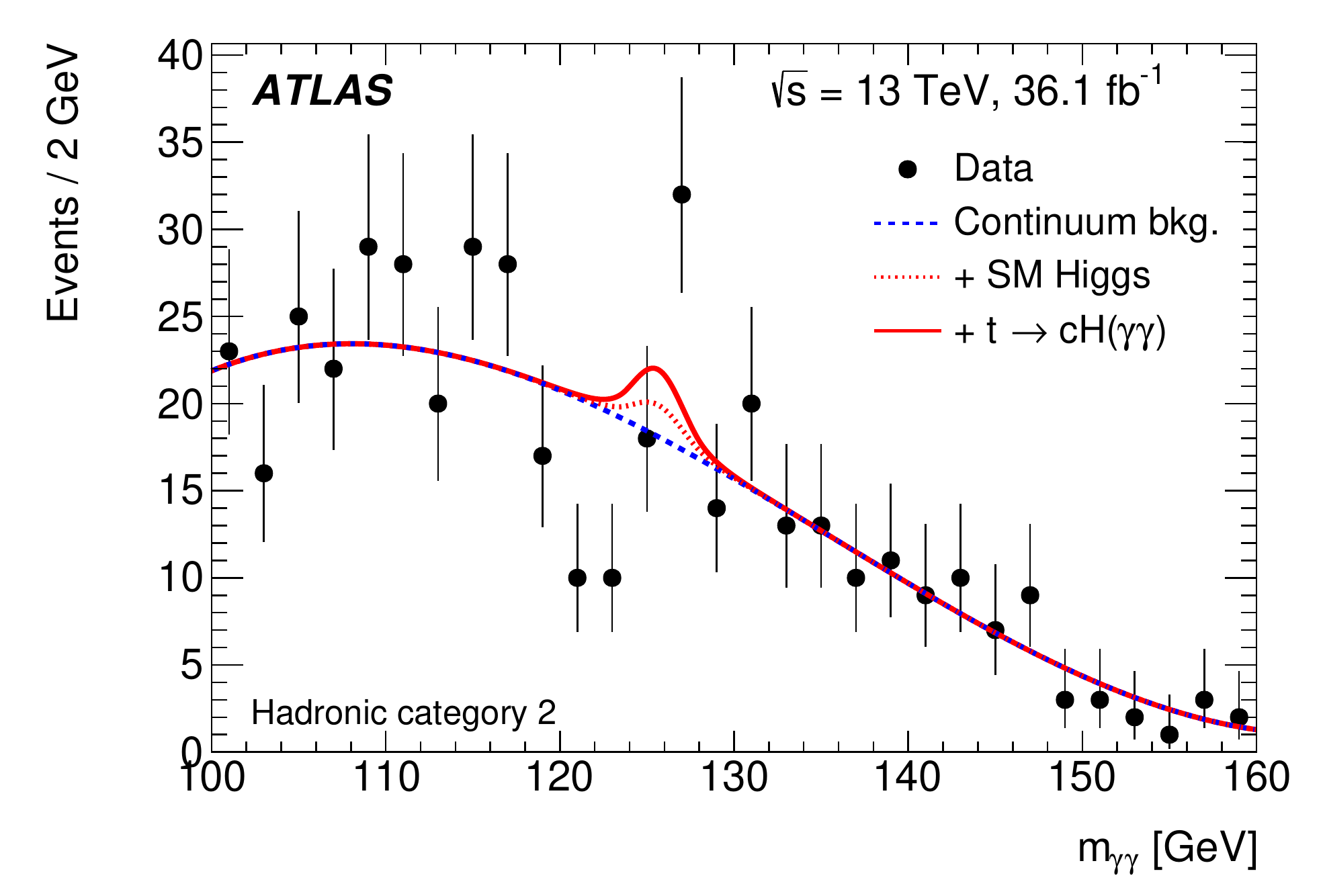}}
  \subfigure[]{\includegraphics[width=.45\textwidth]{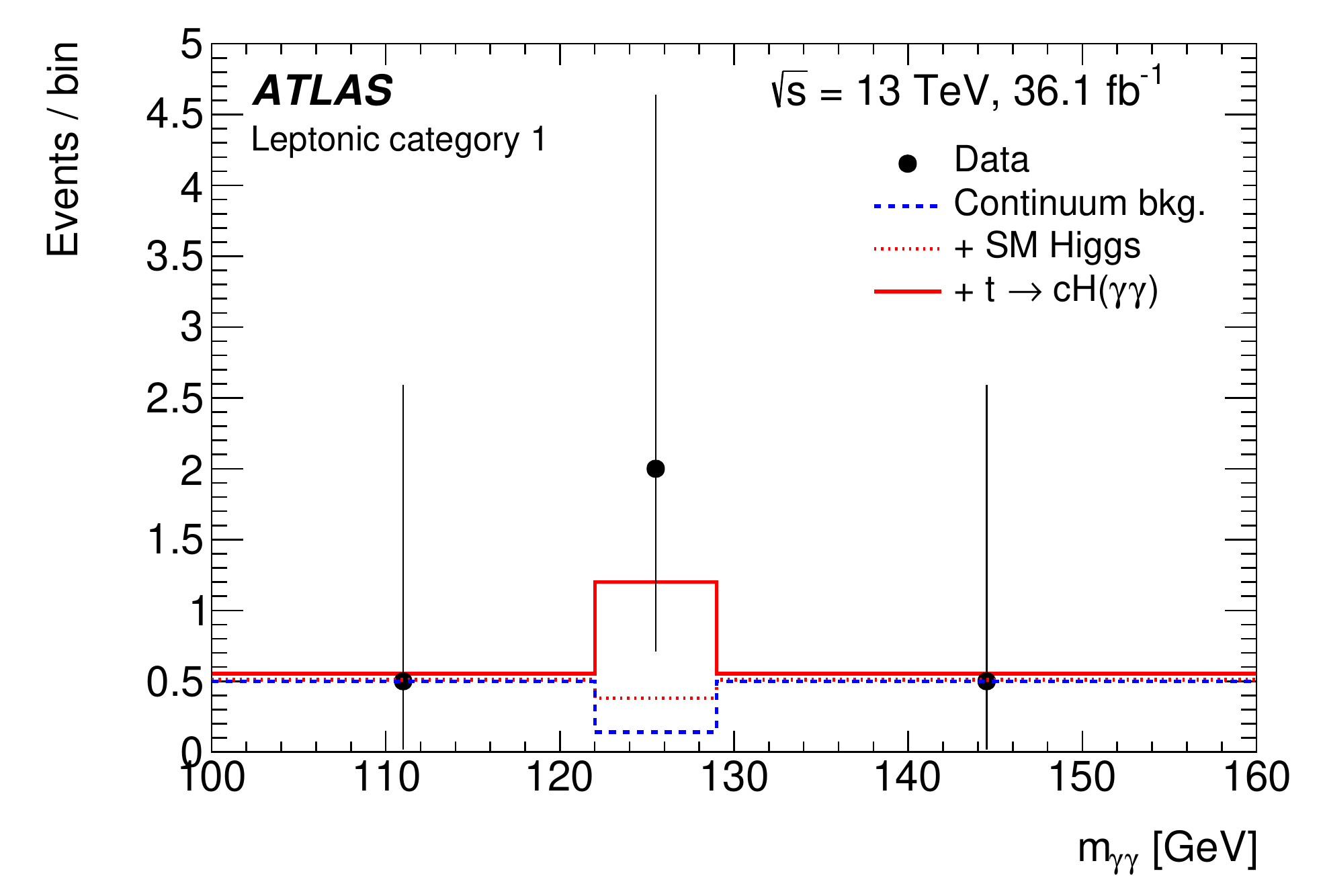}}
  \subfigure[]{\includegraphics[width=.45\textwidth]{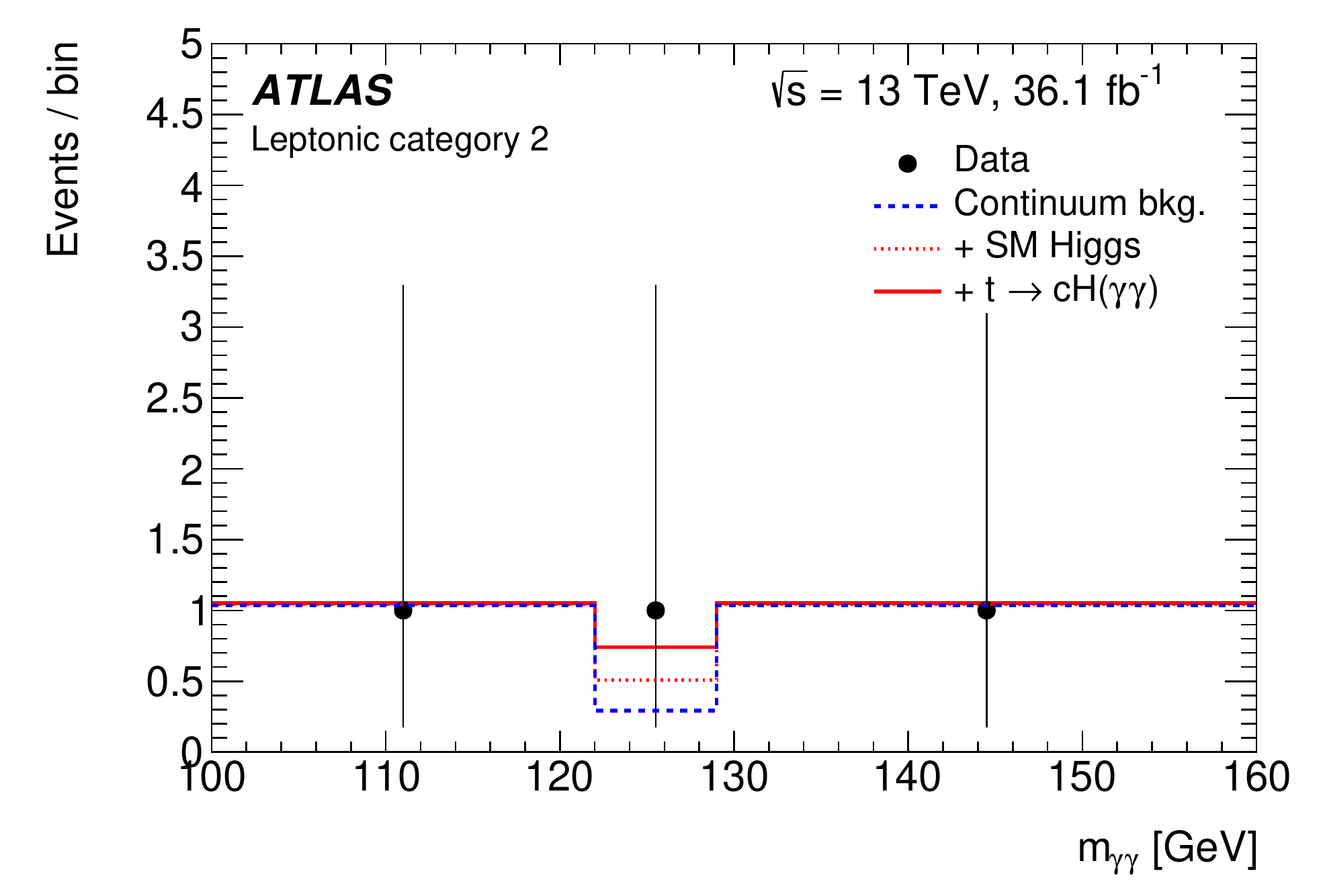}}
\end{center}
\vspace{-0.5cm}
\caption{Distributions of \mgg\ for the selected sample in the (a) hadronic category 1, (b) hadronic category 2, (c) leptonic category 1 and (d) leptonic category 2 channels. The result of fitting the data with the sum (full line)  of a signal component with the mass of the Higgs boson fixed to $m_H = 125.09\ \gev$, a continuum background component (dashed line) and the SM Higgs boson contribution (difference between the dotted and dashed lines) is superimposed. The leptonic categories have only two bins: the seven-\gev-wide SR and the CR (see text). The CR region bin extends from the signal region to both higher and lower masses; the content of the CR is shared equally between the low-mass part (100 \gev\   to 122 \gev) and the high-mass part (129 \gev\  to 160 \gev) of the CR bin.
}
\label{fig:Fit}
\end{figure}

The small excesses observed in leptonic category 1 and hadronic category 2 result in minima of their $q_{\cal B}$ at positive values of ${\cal B}$. On the contrary the tight hadronic selection, where a deficit is observed, pulls  ${\cal B}$ to (unphysical) negative values. The compatibility of the four channels is about 2.3 standard deviations. 
\par The mass distributions corresponding to the result of the combined fit are illustrated for the hadronic and leptonic selections in Figure~\ref{fig:Fit}.
The result of fitting the data with the sum of a signal 
component and a background component (dashed), described by a third-order 
polynomial for the hadronic selection, is superimposed. The small contribution from SM Higgs boson production, included in the fit, is also shown (difference between 
the dotted and dashed lines). For the leptonic channels, the regions $\mgg < 122\ \gev$ and $\mgg > 129\ \gev$ are used as a single-bin control region to estimate the background in the one-bin signal region. For presentation in  Figure~\ref{fig:Fit}, the content of the single-bin CR is shared equally between its low mass part and its high mass part.

\begin{figure}[h!]
\begin{center}
    \includegraphics[width=.5\textwidth]{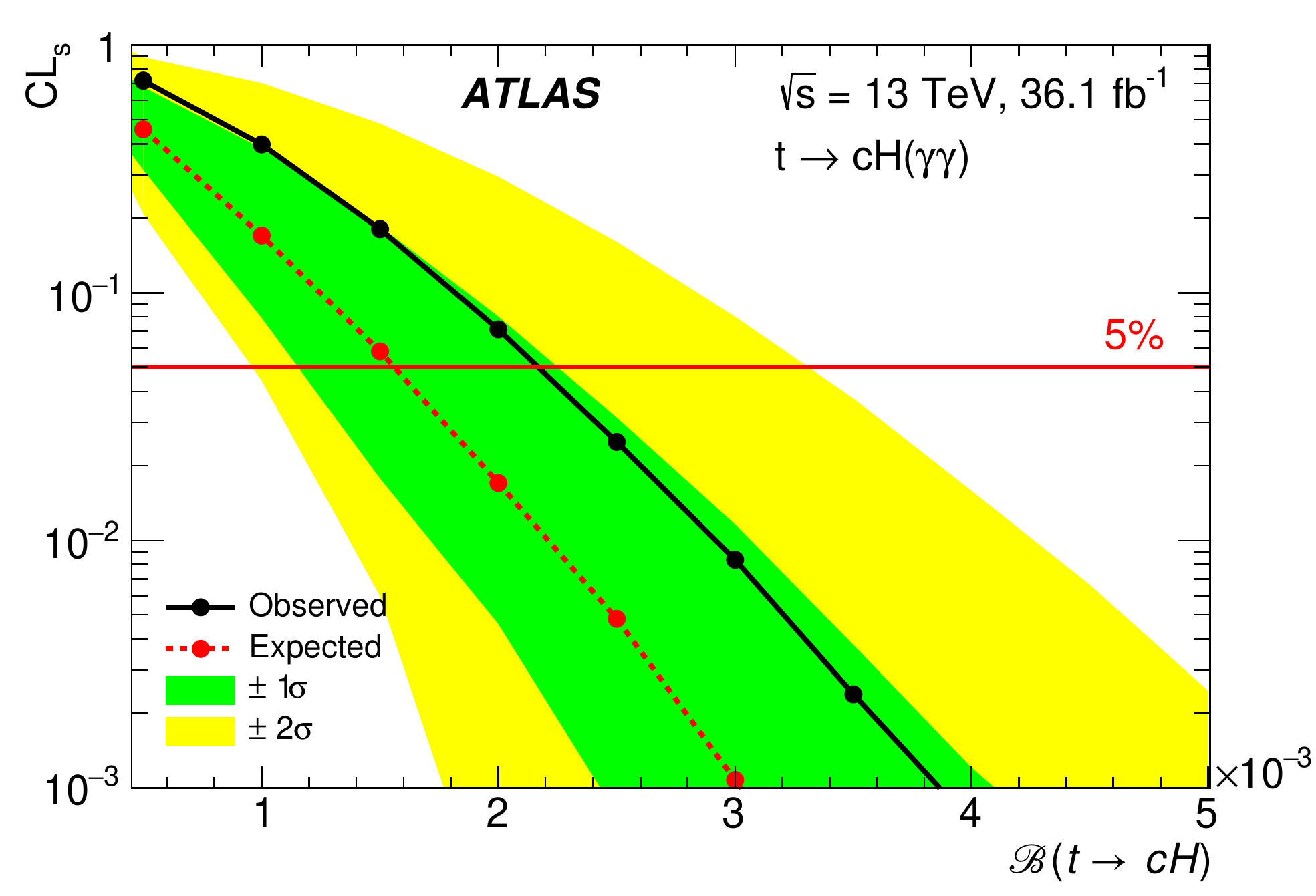}
\end{center}
\vspace{-0.5cm}
\caption{Evolution of the signal confidence level, CL$_{\mathrm{s}}$, as a function of the $t\to cH$ branching ratio ${\cal B}$ for the observation (full line) and the expectation
  in the absence of signal (dashed line). The bands at $\pm 1 \sigma$ and $\pm 2 \sigma$ around the expected curve are also shown.}
\label{fig:olim}
\end{figure}

\par Large-sample pseudo-experiments (which take into account the contribution of the SM Higgs boson production) are used to set the limit on ${\cal B}$.
The evolution of the signal confidence level CL$_{\mathrm{s}}$~\cite{Read:2002hq} as a function of ${\cal B}$, computed from these pseudo-experiments,
is shown in Figure~\ref{fig:olim}, where the observed result ${\cal B}<2.2 \times 10^{-3}$ at the 95\% CL is compared to the expected one in the absence of signal,
${\cal B} < 1.6 \times 10^{-3}$. The limit derived from these pseudo-experiments is close to the value obtained, in the asymptotic approximation, from the intersection of the combined
$q_{\cal B}$ curve with the line at 3.84, marking the 95\% confidence level. It is dominated by statistical uncertainties, as shown in Figure~\ref{fig:2dnll}. 

\par The acceptance of the $t \to uH$ decays is about 8\% lower than for $t \to cH$ in the four analysed channels.
The higher acceptance for $t \to cH$ is mostly driven by the
additional $b$-tagging efficiency given by the charm quark as opposed to the up quark. The observed limit for $t \to uH$ is $2.4\times 10^{-3}$ and the expected limit is $1.7\times 10^{-3}$, both at the 95\% CL. 
\par These limits on ${\cal B}$ can be translated to limits on the off-diagonal Yukawa coupling via the relation 
$$\lambda_{tqH} = (1.92\pm0.02)\times\sqrt{{\cal B}},$$
where the mass of the light quark is neglected~\cite{Aad:2014dya}. The $\lambda_{tqH}$ coupling corresponds to the sum in quadrature of the
couplings relative to the two possible chirality combinations of the quark fields, $\lambda_{tqH} \equiv \sqrt{|\lambda_{t_{\tiny \rm L}q_{\tiny \rm R}}|^2+|\lambda_{q_{\tiny \rm L}t_{\tiny \rm R}}|^2}$~\cite{Harnik:2012pb}.
The observed (expected) limits are
$\lambda_{tcH} < 0.090\ (0.077)$ and $\lambda_{tuH} < 0.094\ (0.079)$ at the 95\% CL.
As the analysis does not 
distinguish between the two channels, the limit can be written as: $\sqrt{\lambda_{tcH}^{2} + 0.92 \lambda_{tuH}^{2}} < 0.090$, where the factor 0.92 is due to the difference in acceptance between the two modes.
With this limit ATLAS reaches the sensitivity region where an observation is possible according to models predicting the largest yields (see Section~\ref{sec:introduction} and Ref.~\cite{Agashe:2013hma}).

\section{Conclusions} 
\par The FCNC decay of a top quark into a lighter up-type quark ($q=c,u$) and a Higgs boson, $t \rightarrow qH$, followed by the decay $H\rightarrow\gamma\gamma$, has been searched for in a data set  of 36.1 \ifb\ of 13 \tev\ proton--proton collisions recorded by the ATLAS experiment at the LHC. The analysis selects events with one top quark decaying into $bW$ (SM decay) and the other into $qH$. It is split into two final states, targeting the decay of the $W$ boson from the SM top quark decay either in the hadronic mode or in the leptonic mode. 

Exploiting the diphoton invariant mass distributions, a sideband technique is used to constrain the background to the sought-for signal. Taking into account the contribution of the SM Higgs boson production, an expected upper limit on the $t\rightarrow cH$ 
decay branching ratio in the absence of signal of $1.6\times 10^{-3}$ is estimated. No statistically significant excess is observed 
in the data, and a limit of $2.2\times 10^{-3}$ is set at the 95\%  CL for $m_H = 125.09\ \gev$. From this limit, an upper limit on the $\lambda_{tcH}$ coupling of 0.090 is obtained.
As the analysis is almost equally sensitive to the $t\rightarrow uH$ mode, the limit obtained can more generally be expressed as $\sqrt{\lambda_{tcH}^{2} + 0.92 \lambda_{tuH}^{2}} < 0.090$.

\section*{Acknowledgements}

We thank CERN for the very successful operation of the LHC, as well as the
support staff from our institutions without whom ATLAS could not be
operated efficiently.

We acknowledge the support of ANPCyT, Argentina; YerPhI, Armenia; ARC, Australia; BMWFW and FWF, Austria; ANAS, Azerbaijan; SSTC, Belarus; CNPq and FAPESP, Brazil; NSERC, NRC and CFI, Canada; CERN; CONICYT, Chile; CAS, MOST and NSFC, China; COLCIENCIAS, Colombia; MSMT CR, MPO CR and VSC CR, Czech Republic; DNRF and DNSRC, Denmark; IN2P3-CNRS, CEA-DSM/IRFU, France; SRNSF, Georgia; BMBF, HGF, and MPG, Germany; GSRT, Greece; RGC, Hong Kong SAR, China; ISF, I-CORE and Benoziyo Center, Israel; INFN, Italy; MEXT and JSPS, Japan; CNRST, Morocco; NWO, Netherlands; RCN, Norway; MNiSW and NCN, Poland; FCT, Portugal; MNE/IFA, Romania; MES of Russia and NRC KI, Russian Federation; JINR; MESTD, Serbia; MSSR, Slovakia; ARRS and MIZ\v{S}, Slovenia; DST/NRF, South Africa; MINECO, Spain; SRC and Wallenberg Foundation, Sweden; SERI, SNSF and Cantons of Bern and Geneva, Switzerland; MOST, Taiwan; TAEK, Turkey; STFC, United Kingdom; DOE and NSF, United States of America. In addition, individual groups and members have received support from BCKDF, the Canada Council, CANARIE, CRC, Compute Canada, FQRNT, and the Ontario Innovation Trust, Canada; EPLANET, ERC, ERDF, FP7, Horizon 2020 and Marie Sk{\l}odowska-Curie Actions, European Union; Investissements d'Avenir Labex and Idex, ANR, R{\'e}gion Auvergne and Fondation Partager le Savoir, France; DFG and AvH Foundation, Germany; Herakleitos, Thales and Aristeia programmes co-financed by EU-ESF and the Greek NSRF; BSF, GIF and Minerva, Israel; BRF, Norway; CERCA Programme Generalitat de Catalunya, Generalitat Valenciana, Spain; the Royal Society and Leverhulme Trust, United Kingdom.

The crucial computing support from all WLCG partners is acknowledged gratefully, in particular from CERN, the ATLAS Tier-1 facilities at TRIUMF (Canada), NDGF (Denmark, Norway, Sweden), CC-IN2P3 (France), KIT/GridKA (Germany), INFN-CNAF (Italy), NL-T1 (Netherlands), PIC (Spain), ASGC (Taiwan), RAL (UK) and BNL (USA), the Tier-2 facilities worldwide and large non-WLCG resource providers. Major contributors of computing resources are listed in Ref.~\cite{ATL-GEN-PUB-2016-002}.

\clearpage
\printbibliography

\newpage
\begin{flushleft}
{\Large The ATLAS Collaboration}

\bigskip

M.~Aaboud$^\textrm{\scriptsize 137d}$,
G.~Aad$^\textrm{\scriptsize 88}$,
B.~Abbott$^\textrm{\scriptsize 115}$,
O.~Abdinov$^\textrm{\scriptsize 12}$$^{,*}$,
B.~Abeloos$^\textrm{\scriptsize 119}$,
S.H.~Abidi$^\textrm{\scriptsize 161}$,
O.S.~AbouZeid$^\textrm{\scriptsize 139}$,
N.L.~Abraham$^\textrm{\scriptsize 151}$,
H.~Abramowicz$^\textrm{\scriptsize 155}$,
H.~Abreu$^\textrm{\scriptsize 154}$
% ...
\bigskip
\\
% ...
$^{12}$ Institute of Physics, Azerbaijan Academy of Sciences, Baku, Azerbaijan\\
% ...
$^{88}$ CPPM, Aix-Marseille Universit{\'e} and CNRS/IN2P3, Marseille, France\\
% ...
$^{115}$ Homer L. Dodge Department of Physics and Astronomy, University of Oklahoma, Norman OK, United States of America\\
% ...
$^{119}$ LAL, Univ. Paris-Sud, CNRS/IN2P3, Universit{\'e} Paris-Saclay, Orsay, France\\
% ...
$^{137}$ $^{(a)}$ Facult{\'e} des Sciences Ain Chock, R{\'e}seau Universitaire de Physique des Hautes Energies - Universit{\'e} Hassan II, Casablanca; $^{(b)}$ Centre National de l'Energie des Sciences Techniques Nucleaires, Rabat; $^{(c)}$ Facult{\'e} des Sciences Semlalia, Universit{\'e} Cadi Ayyad, LPHEA-Marrakech; $^{(d)}$ Facult{\'e} des Sciences, Universit{\'e} Mohamed Premier and LPTPM, Oujda; $^{(e)}$ Facult{\'e} des sciences, Universit{\'e} Mohammed V, Rabat, Morocco\\
% ...
$^{139}$ Santa Cruz Institute for Particle Physics, University of California Santa Cruz, Santa Cruz CA, United States of America\\
% ...
$^{151}$ Department of Physics and Astronomy, University of Sussex, Brighton, United Kingdom\\
% ...
$^{154}$ Department of Physics, Technion: Israel Institute of Technology, Haifa, Israel\\
$^{155}$ Raymond and Beverly Sackler School of Physics and Astronomy, Tel Aviv University, Tel Aviv, Israel\\
% ...
$^{161}$ Department of Physics, University of Toronto, Toronto ON, Canada\\
% ...
$^{a}$ Also at Department of Physics, King's College London, London, United Kingdom\\
$^{b}$ Also at Institute of Physics, Azerbaijan Academy of Sciences, Baku, Azerbaijan\\
$^{c}$ Also at Novosibirsk State University, Novosibirsk, Russia\\
$^{d}$ Also at TRIUMF, Vancouver BC, Canada\\
$^{e}$ Also at Department of Physics {\&} Astronomy, University of Louisville, Louisville, KY, United States of America\\
% ...
$^{*}$ Deceased
\end{flushleft}


\end{document}